\documentclass[preprint2]{aastex}

\def\etal{et~al.}

\def\vi{\ifmmode(V{-}I)\else$(V{-}I)$\fi}
\def\viz{\ifmmode(V{-}I)_0\else$(V{-}I)_0$\fi}

\def\gz{\ifmmode(g_{475}{-}z_{850})\else$(g_{475}{-}z_{850})$\fi}
\def\gzz{\ifmmode(g_{475}{-}z_{850})_0\else$(g_{475}{-}z_{850})_0$\fi}
\newcommand\lta{\mathrel{\rlap{\lower 3pt\hbox{$\mathchar"218$}}
     \raise 2.0pt\hbox{$\mathchar"13C$}}}
\newcommand\gta{\mathrel{\rlap{\lower 3pt\hbox{$\mathchar"218$}}
     \raise 2.0pt\hbox{$\mathchar"13E$}}}
\def\mbari{\ifmmode\overline{m}_I\else$\overline{m}_I$\fi}
\def\mbarz{\ifmmode\overline{m}_z\else$\overline{m}_z$\fi}
\def\mbar{\ifmmode\overline{m}\else$\overline{m}$\fi}
\def\Mbar{\ifmmode\overline{M}\else$\overline{M}$\fi}
\def\Mbarz{\ifmmode\overline{M_z}\else$\overline{M}_z$\fi}

\shorttitle{CL0910}
\shortauthors{Mei et al.}

\begin{document}


\title{Evolution of the Color-Magnitude Relation in High-Redshift Clusters:
Blue Early-Type Galaxies and Red Pairs in RDCS~J0910+5422}

\author{S. Mei\altaffilmark{1},
J. P. Blakeslee\altaffilmark{1}, 
S. A. Stanford\altaffilmark{2,3},
B. P.~Holden \altaffilmark{4}, 
P. Rosati\altaffilmark{6},
V. Strazzullo\altaffilmark{20,6},
N.~Homeier\altaffilmark{1},
M. Postman\altaffilmark{1,5},
M. Franx\altaffilmark{12}, 
A. Rettura\altaffilmark{6},
H. Ford\altaffilmark{1},
G. D. Illingworth \altaffilmark{4},
S. Ettori\altaffilmark{19},
R.J.~Bouwens\altaffilmark{4},
R. Demarco\altaffilmark{1},
A.R. Martel\altaffilmark{1},
M. Clampin\altaffilmark{5},
G.F. Hartig\altaffilmark{5},
P. Eisenhardt\altaffilmark{7},
D.R.~Ardila\altaffilmark{1},
F. Bartko\altaffilmark{8}, 
N. Ben\'{\i}tez\altaffilmark{18},
L.D. Bradley\altaffilmark{1},
T.J. Broadhurst\altaffilmark{9},
R.A. Brown\altaffilmark{5},
C.J. Burrows\altaffilmark{5},
E.S. Cheng\altaffilmark{10},
N.J.G. Cross\altaffilmark{17},
P.D. Feldman\altaffilmark{1},
D.A. Golimowski\altaffilmark{1},
T. Goto\altaffilmark{1},
C. Gronwall\altaffilmark{13},
L. Infante\altaffilmark{14}
R.A. Kimble\altaffilmark{11},
J.E. Krist\altaffilmark{5},
M.P. Lesser\altaffilmark{15},
F. Menanteau\altaffilmark{1},
G.R. Meurer\altaffilmark{1},
G.K. Miley\altaffilmark{12},
V. Motta\altaffilmark{14},
M. Sirianni\altaffilmark{5}, 
W.B. Sparks\altaffilmark{5}, 
H.D. Tran\altaffilmark{16}, 
Z.I.~Tsvetanov\altaffilmark{1},   
R.L. White\altaffilmark{5},
\& W. Zheng\altaffilmark{1}}
\altaffiltext{1}{Dept.\ of Physics \&Astronomy, Johns Hopkins University, Baltimore, MD 21218; smei@pha.jhu.edu}
\altaffiltext{2}{Department of Physics, University of California, Davis, CA 94516}
\altaffiltext{3}{Institute of Geophysics and Planetary Physics, Lawrence Livermore National Lab, Livermore, CA 94551}
\altaffiltext{4}{Lick Observatory, University of California, Santa Cruz, CA 95064}
\altaffiltext{5}{Space Telescope Science Institute, 3700 San Martin Drive, Baltimore, MD 21218}
\altaffiltext{6}{European Southern Observatory, Karl-Schwarzschild-Str. 2, D-85748 Garching, Germany}
\altaffiltext{7}{Jet Propulsion Laboratory, CalTech, 4800 Oak Grove Drive, Pasadena, CA 91125}
\altaffiltext{8}{Bartko Science \& Technology, 14520 Akron Street, 
Brighton, CO 80602.}    
\altaffiltext{9}{School of Physics and Astronomy, Tel Aviv University, Tel Aviv 69978, Israel}
\altaffiltext{10}{Conceptual Analytics, LLC, 8209 Woburn Abbey Road, Glenn Dale, MD 20769.}
\altaffiltext{11}{NASA Goddard Space Flight Center, Code 681, Greenbelt, MD 20771.}
\altaffiltext{12}{Leiden Observatory, Postbus 9513, 2300 RA Leiden,
Netherlands.}
\altaffiltext{13}{Dept.\ of Astronomy \& Astrophysics, 
Penn State University, University Park, PA 16802.}
\altaffiltext{14}{Dept.\ de Astronom\'{\i}a y Astrof\'{\i}sica,
Pontificia Universidad Cat\'{\o}lica, Casilla 306, Santiago
22, Chile.} 
\altaffiltext{15}{Steward Observatory, University of Arizona, Tucson,
AZ 85721.} 
\altaffiltext{16}{W. M. Keck Observatory, 65-1120 Mamalahoa Hwy., 
Kamuela, HI 96743}
\altaffiltext{17}{Royal Observatory Edinburgh, Blackford Hill, Edinburgh, EH9 3HJ, UK}
\altaffiltext{18}{Instituto de Astrof\'\i sica de Andaluc\'\i a (CSIC), Camino Bajo de Hu\'etor 50, Granada 18008, Spain }
\altaffiltext{19}{INAF - Osservatorio Astronomico, via Ranzani 1, 40127 Bologna, Italy }
\altaffiltext{20}{Dipartimento di Scienze Fisiche, Universit\`a Federico II, I-80126 Napoli,Italy}
\begin{abstract}
The color-magnitude relation has been determined for the RDCS~J0910+5422 cluster of galaxies at redshift z = 1.106.  Cluster members
were selected from HST Advanced Camera for Surveys (ACS) images,
combined with ground--based near--IR imaging and optical
spectroscopy.  Postman et al. (2005) morphological classifications
were used to identify the early-type galaxies.  

The
observed early--type color--magnitude relation (CMR) in $(i_{775} -
z_{850})$ versus $z_{850}$ shows an intrinsic scatter in color of $0.060 \pm
0.009$~mag, within $1\arcmin$ from the cluster X--ray emission center. 
Both the ellipticals and the S0s show small scatter about the CMR of $0.042 \pm 0.010$~mag and
$0.044 \pm 0.020$~mag, respectively.  From the scatter about the CMR,
a mean luminosity--weighted age $\overline t > 3.3$~Gyr ($z_f > $~3)
is derived for the elliptical galaxies, assuming a simple stellar population modeling (single burst solar metallicity).
This is consistent with a previous study of the cluster RDCS\,1252.9-292 at $z{=}1.24$ (Blakeslee et al.).

Strikingly, the S0 galaxies in RDCS~J0910+5422 are
systematically bluer in $(i_{775}{-}z_{850})$ by $0.07 \pm 0.02$~mag,
with respect to the ellipticals.
The blue S0s are predominantly elongated in shape; the distribution of 
their ellipticities is inconsistent with a population of axisymmetric
disk galaxies viewed at random orientations, suggesting either that they are
intrinsically prolate or there is some orientation bias 
in the S0 classification. 
The ellipticity distribution as
a function of color indicates that the face-on S0s in this particular cluster have likely 
been classified as elliptical.  
Thus, if anything, the offset in color
between the elliptical and S0 populations may be even more significant.

The color offset between S0 and E corresponds to an age difference of $\approx$~1~Gyr, for a single-burst
solar metallicity model. Alternatively, it could be the result of
a different star formation history; a solar metallicity model 
with an exponential decay in star formation will reproduce the offset for an age of 3.5 Gyr, i.e. the S0s have evolved gradually from star forming
progenitors.
The color offset could also be reproduced by a factor of $\sim\,$2 difference
in metallicity, but the two populations would each need to have very small
scatter in metallicity to reproduce the small scatter in color.

The early--type population in this cluster appears to be still forming.
The blue early-type disk galaxies in RDCS~J0910+5422 likely represent the direct progenitors of the more evolved
S0s that follow the same red sequence as ellipticals in other clusters. 

Thirteen red galaxy pairs are observed and the galaxies associated in pairs
constitute $\sim$40\% of the CMR galaxies in this cluster. This finding
is consistent with the conclusions of van Dokkum and Tran et al. that most of 
the early--type galaxies grew from passive red mergers.

 \end{abstract}

\section{Introduction}

The Advanced Camera for Surveys (ACS; Ford et al. 2002), by virtue
of its high spatial resolution and sensitivity,
allows us to study galaxy clusters in great detail up to
redshifts of unity and beyond. At these redshifts, galaxy
clusters are still assembling and galaxies are evolving towards the
populations that we observe today.    Recent results
from our ACS Intermediate Redshift Cluster Survey (Blakeslee et al. 2003a; Lidman et
al. 2004; Demarco et al. 2005; Goto et al. 2005; Holden et al. 2005a;
Holden et al. 2005b; Homeier et al. 2005; Postman et al. 2005) have
shown that galaxy clusters at redshift around unity show many
similarities with local clusters, in terms of galaxy populations and
their distribution, but also significant differences in galaxy
morphology, ellipticity, and mass--luminosity ratios. 
The strongest evolution observed in the early--type population is
 a deficit of a S0 population in this sample when compared to lower 
redshift samples (Postman et al. 2005). This would give evidence 
that the formation of the S0 population is still under way in clusters 
at redshift unity. 

One of the most
striking similarities is that the tight relation between early-type
galaxy colors and luminosities that applies locally
(the color--magnitude relation; CMR) is already in place at redshifts as
high as $z \sim 1.3$ (e.g. Stanford et al.\ 1997; Mullis et al.\ 2005). The
CMR in local samples of galaxy clusters presents universal properties,
in terms of scatter and zero point (Bower et al. 1992; van Dokkum et
al. 1998, Hogg et al. 2004; L\'opez--Cruz et al. 2004; Bell et
al. 2004; Bernardi et al. 2005; McIntosh et al. 2005) that evolve back
in time in agreement with passively evolving models 
(Ellis et al.\ 1997; Stanford, Eisenhardt, \& Dickinson 1998; van
Dokkum et al. 2000, 2001; Blakeslee et al. 2003a; Holden et al. 2004;
De Lucia et al. 2004; Blakeslee et al. 2005). ACS enables accurate measurement of the scatter
around the CMR, with enough precision to seriously constrain galaxy formation age, 
which is impossible to obtain from ground--based data (see for example Holden et al. 2004). The measurement of the CMR scatter of the first cluster in our ACS
cluster survey, RXJ1252.9-292, permitted us to constrain the mean luminosity--weighted age for the ellipticals to be $> 2.6$~Gyr ($z > 2.7$) (Blakeslee et
al. 2003a), based on simple modeling.  In this paper,
we extend the results obtained in Blakeslee et
al. (2003a) to RXJ~0910+5422.

RXJ~0910+5422 is part of the ACS cluster survey (guaranteed time
observation, GTO, program \#9919), that includes eight
clusters in the redshift range at $0.8 < z < 1.3$, selected in
the X--ray, optical and near--IR (Ford et al.\ 2004). 
RXJ~0910+5422 was selected from the ROSAT
Deep Cluster Survey (Rosati et al. 1998) and confirmed with near-IR and
spectroscopic observations by Stanford et al. (2002). Extensive
followup spectroscopy at the Keck Observatory has been carried out in
a magnitude limited sample reaching $K_s = 20.0$~mag in the central
$3$~arcmin (Stanford et al.\ 2005; in preparation).  The mean redshift of the cluster
was measured to be $z=1.106$ (Stanford et al. 2002).  In this paper,
 we combine ACS imaging
with ground--based spectroscopy and near--IR imaging to constrain galaxy
ages and formation histories from the study of their color--magnitude relation. 
We discuss the properties of the elliptical (E) and lenticular (S0) 
populations separately in the light of simple galaxy formation scenarios.

\section{Observations}

RXJ~0910+5422 was observed in March 2004 with the ACS WFC (Wide
Field Camera) in the F775W ($i_{775}$) and F850LP
($z_{850}$) bandpasses, with total exposure times of 6840~s and 11440~s,
respectively. The ACS WFC scale is 0.05\arcsec/pixel, and its field of
view is $210\arcsec \times 204\arcsec$. The APSIS pipeline (Blakeslee
et al. 2003b), with a {\it Lanczos3} interpolation kernel, was used
for processing the images.  The ACS photometric zero--points (AB system)
are 25.654~mag and 24.862~mag in $i_{775}$ and $z_{850}$, respectively
(Sirianni et al. 2005). A Galactic reddening of $E(B-V)=0.019$ towards
RXJ~0910+5422 was adopted (Schlegel et al. 1998), with $A_{i775}=0.039$
and $A_{z850}=0.029$ (Sirianni et al. 2005).  The ACS WFC field covers
an area that at the redshift of this cluster, $z=1.106$, corresponds
to $\approx 1 Mpc^2$ in the WMAP cosmology (Spergel et al. (2003):
$\Omega_m =0.27$, $\Omega_{\Lambda} =0.73$, $h=0.71$, adopted as our
standard cosmology hereafter). Fig.~\ref{cluster} shows the ACS color
image with X--ray contours from Chandra ACIS (Advanced CCD Imaging
Spectrometer) data that have been adaptively smoothed (Stanford et
al. 2002).
Near-IR $JK_s$ and optical $i$-band images were obtained at Palomar
Observatory as described in detail by Stanford et al. (2002).  

Optical spectroscopy of galaxies in RXJ~0910+5422 was obtained using the Low
Resolution Imaging Spectrometer (LRIS; Oke et al. 1995) on the Keck 1
and 2 telescopes (Stanford et al. 2005; in preparation).  
Our typical errors in redshift correspond to errors in velocity between
100 and 300~km/s. 
Objects for spectroscopy were chosen
initially from the catalog of objects with $K_s < 20.0$~mag (Vega
magnitudes) within the IR imaging area; outside of this area objects
were chosen with $i > 21$~mag from the $i$-band 
image to fill out masks. Our final sample included 66\% of the
objects with $K_s < 20.0$~mag.
Spectra were obtained using the 400 lines mm$^{-1}$ grating for all
runs except for the initial two discovery masks as reported in
Stanford et al.\ (2002).  Nine more masks were observed using LRIS
during four runs between January 2001 and February 2003.  Usually each
mask was observed in a series of four 1800~s exposures, with small
spatial offsets along the long axis of the slits.  On average, the
 seeing was 0.9~$\arcsec$.  The blue side data were generally not used since the rest
frame wavelengths probed at $z = 1.1$ fall far to the blue of the
spectral features of interest for galaxies in the cluster.
In total, 149 redshifts were obtained.

The slit mask data were separated into slitlet spectra and then
reduced using standard long-slit techniques.   A fringe frame was
constructed for each exposure from neighboring exposures, each offset from
the previous by 3$\arcsec$, in an
observing sequence for each mask,
 and then subtracted from each
exposure to greatly reduce fringing in the red.  The exposures for
each slitlet were reduced separately and then co-added.
One-dimensional spectra were extracted for each targeted object, as
well as the occasional serendipitous source.  Wavelength calibration
of the 1-D spectra was obtained from arc lamp exposures taken
immediately after the object exposures.  A relative flux calibration
was obtained from long-slit observations of the standard stars HZ44,
G191B2B, and Feige 67 (Massey \& Gronwall 1990).

\section{Object selection and photometry}

SExtractor (Bertin \& Arnouts 1996)
was used to find objects  in the $i_{775}$ and $z_{850}$ images and measure their magnitudes.
Threshold and deblending settings were used  as in Ben\'{\i}tez et al. (2004). 
Although we have extensive spectroscopy, the ACS imaging 
reaches considerably deeper along the cluster luminosity function.  Thus, we have
chosen to use colors ($i_{775}$ - $z_{850}$) and $(J-K_s)$ to isolate
a set of probable cluster members.
In Fig.~\ref{colage}, the ($i_{775}$ - $z_{850}$) and $(J-K_s)$
colors are shown as a function of galaxy age, using BC03 stellar population models,
redshifted to $z{=}1.106$. Early--type cluster members would have ages 
of at least 0.5~Gyr, corresponding to $(i_{775} - z_{850}) >$0.8~mag and
$(J-K_s) > 1.45$~mag.
At first, we give a larger color margin and select as potential cluster members all morphologically-classified early-type galaxies  with $0.5 <(i_{775} -
z_{850}) < 1.2$~mag and $(J-K_s) > 1.45$~mag, down to $z_{850} = 24$~mag (the limiting magnitude
of Postman et al. (2005) morphological classification, that included all clusters
in our sample at redshift unity). Our results in this paper are based on this morphological classification and a detailed discussion of the uncertainties in this classification can be found in that work.
 This selected sample
includes 38 galaxies within the ACS field.  

Our final colors
were measured within galaxy effective radii ($R_e$), to avoid biases
due to galaxy internal gradients, following the approach in Blakeslee
et al. (2003a) and van Dokkum et al. (1998, 2000).  $R_e$ values were
derived with the program GALFIT (Peng et al. 2002), constraining the
{\it Sersic} index $n \le 4$ (as in Blakeslee et al. 2003a). 
To remove differential blurring effects
(the PSF is $\sim10\%$ broader in the $z_{850}$ band) each galaxy image in
both $i_{775}$ and $z_{850}$ was deconvolved using the CLEAN algorithm
(H{\"o}gbom et al. 1974).  The $(i_{775} - z_{850})$ colors were
measured on the deconvolved images within a circular aperture of radius equal to
$R_e$, or 3 pixels, whichever is larger.  Our median $R_e$ is $\approx$~5.5~pixels
($\approx$~13kpc at z=1.106).
Our final results do not change (within the uncertainties) if the 
 effective radii are calculated via a two component 
(Sersic bulge + exponential disk) surface brightness decomposition 
technique using GIM2D (Marleau \&  
Simard 1998; Rettura et al., in preparation), that permits us to better  fit 
the galaxy light profile.

The photometric uncertainties
due to flat fielding, PSF variations, and the pixel-to-pixel
correlation for ACS (Sirianni et al. 2005) were estimated by measuring
the standard deviation of photometry in the background for circular
apertures in the range of the measured effective radii. These
photometric errors were added in quadrature to the Poisson
uncertainties in the measured fluxes for each object.  The derived
errors in the colors are between 0.01 and 0.03~mag down to $z_{850} =
24$~mag. SExtractor MAG\_AUTO were used for the $z_{850}$ magnitude in 
the color--magnitude relation; these are fairly robust, though may
systematically miss a small fraction of the light (Ben\'{\i}tez et al.\ 2004).

We finally color--selected 34 early--type (E, S0 and S0/a) galaxies with $0.8
<(i_{775} - z_{850}) < 1.1$~mag within 2$\arcmin$ from the cluster center,
taken as the center of the X--ray emission (Stanford et al.\ 2002).
Images of the color--selected galaxies are shown in 
Fig~\ref{epostage}, Fig~\ref{sopostage} and Fig~\ref{sapostage}.
Moreover, there are late--type galaxies with
luminosities that are similar to the red--sequence bright
early--type galaxies (Fig.~\ref{spipostage}).
Of the 34 color--selected galaxies, 15 are spectroscopically confirmed cluster members,
one (S0/a, with magnitude $z_{850} = 24.2$~mag) is a  confirmed non-member,
and the others were not targeted for spectroscopy.  The
selection in $(i_{775} - z_{850})$ at $z=1.1$ therefore appears to be 
robust: only one of the 16 selected galaxies with measured redshifts
is a non-member.
We expect few of the other 18 to be interloper field galaxies.

\section{Color-Magnitude Relation}

The color--magnitude relation for the final color--selected objects
 is shown in Fig~\ref{cmd}.  Red dots
are ellipticals, orange squares and stars are S0 and S0/a galaxies,
respectively.  Smaller black symbols represent early--type galaxies that do
not lie on the red sequence. Small triangles are late-type
galaxies. Boxes are plotted around confirmed cluster members. 
Confirmed interlopers are circled in the figure.
Surprisingly, the two brightest
cluster members are not ellipticals, but S0. The brightest of these
two galaxies lies $\approx$~700~kpc ( $\approx$1.2$\arcmin$) from
the cluster center, and the other bright S0 at $\approx$~300~kpc 
($\approx$~0.6$\arcmin$).  Moreover, there are late--type galaxies with 
luminosities that are similar to the red--sequence bright
early--type galaxies. Two of them lie on the red--sequence and are
confirmed cluster members, at $\approx$~80~kpc from the cluster center
(see also below in the discussion of the color and morphology
distribution as a function of distance from the cluster center).

We fitted the following linear color--magnitude relation to various 
subsamples of the galaxies:
\begin{equation}
i_{775} - z_{850} =  c_0 + Slope  (z_{850} - 23)
\end{equation}
The solid line in Fig~\ref{cmd} is the fit to the
color--magnitude relation for the ellipticals, the black dotted line is
the fit to the CMR for the S0s, and the dashed--dotted line
the fit to the full sample, within 2$\arcmin$ from the cluster center (see discussion below). The
dashed line is the fit to the full sample of early--type galaxies in
RXJ1252.9-292 from Blakeslee et al. (2003a), scaled to this
redshift with BC03 evolved stellar population models, with solar
metallicity and a formation age of 2.6~Gyr (since Blakeslee et al. 2003a
obtains elliptical mean ages ~$>$~2.6~Gyr). The long--dashed vertical line is the magnitude limit
of the morphological classification $z_{850} = 24$~mag. 
The results for different morphological samples are given in Table~\ref{results}. 

A robust linear fit based on Bisquare weights (Tukey's biweight; Press et al. (1992)) has been used to fit the color--magnitude relation. Uncertainties on the parameters were estimated by bootstrapping on 10,000 simulations. The scatter around the fit was estimated from a biweight scale estimator (Beers, Flynn \& Gebhardt 1990), that is insensitive to outliers, in the same set of bootstrap simulations.
The internal color scatter ($\sigma_{int}$) was measured in two ways:
1) to the scatter around the fit, we have subtracted in quadrature the
average uncertainty due to the galaxy color error; and 2) we have
calculated the internal scatter for which the $\chi^2$ of the fit
would be unity.  Both methods give us internal scatters consistent to
within a few 0.001~mag.  All galaxies in this sample lie within 
three sigma from the fit.

The X--ray distribution appears to be very symmetric, and largely confined
within 1\farcm5 from the cluster center. We
calculated the CMR zero point and scatter within 1$\arcmin$ (which
corresponds to a scale of $\approx$~0.5~Mpc at this redshift), and within 
1\farcm5 ($\approx$0.7~Mpc), and within 2$\arcmin$
($\approx$~1~Mpc, the scale
used for the analysis of RXJ1252.9-292).  According to the results in
Table~\ref{results}, the internal color scatter increases when adding
populations between 1$\arcmin$ and 2$\arcmin$, especially for the S0
and S0/a populations, as also observed in local samples (e.g. van Dokkum
et al. 1998), with only a small increase in sample size.
We will therefore focus on the results obtained for color--selected galaxies 
within 1$\arcmin$ from the cluster
center (where 90\% of the color-selected galaxies lie).  The slope of the
elliptical CMR ($-0.033 \pm 0.015$) is slightly 
steeper than the
observed slope in RXJ1252.9-292 ($-0.020 \pm 0.009$), and in Coma when
the latter are shifted to the observed colors at $z \sim 1.1$, using 
non-evolving BC03 stellar population models, but still consistent within the uncertainties.  We do
not find a flatter (with respect to Coma) slope as in Stanford et
al. (2002).  However, the S0 sample shows a much shallower slope
($0.005 \pm 0.023$) than the ellipticals, resulting in a much flatter slope
for ellipticals and S0s together ($-0.024 \pm 0.020$). This can explain why a
shallower slope was found in that work, in which elliptical and
S0s were not separated. The spectroscopically-selected elliptical plus
S0 slope ($-0.010 \pm 0.034$) is also flattened by the S0 population,
while the spectroscopically-selected ellipticals have a slope ($-0.021 \pm 0.046$) similar to that of RXJ1252.9-292 and Coma. All the difference in slope are 
however within the uncertainties and are statistically insignificant.

Using Bruzual \& Charlot (2003; BC03) stellar population models, as in
Blakeslee et al. (2003a), we derive a constraint on the age of the stellar
populations in the galaxies from galaxy colors and the
scatter of the CMR (van Dokkum et al. 2001; Blakeslee et
al. 2003a). Two simple models have been considered and our conclusions will
depend on the chosen models. The first model is a {\it single burst} model,
in which galaxies form in single bursts at random times $t_f$,
between the age of the cluster and the recombination epoch. The second is a
model with {\it constant star formation} in a range of time between $t_1$
and $t_2$, randomly chosen to be between the age of the cluster and
the recombination epoch. Colors for 10,000 galaxies were simulated
with their scatter around the CMR to be
dependent on the burst age.  
In Fig~\ref{scatterage}, we show the simulated scatters as a function
of burst age, with solar, half solar and twice solar
metallicity model. We will assume solar metallicity in what follows.
From the scatter ($\sigma_{int}=0.042\pm
0.010$) in the colors of the galaxies classified as ellipticals, we obtain ages $>2.1$~Gyr
($z >2$), with a mean luminosity--weighted age $\overline t=3.31$~Gyr
($z_f \approx$~3.1), assuming the random single burst model. From the constant star
formation model, we obtain ages $>1.6$~Gyr ($z >1.7$), with a mean
luminosity--weighted age $\overline t=3.26$~Gyr
($z_f \approx$~3). This agrees with the conclusion (e.g. Blakeslee et
al. 2003a; Holden et al. 2004; Lidman et al. 2004; De Lucia et
al. 2005) that the elliptical population in clusters of galaxies
formed at $z_f >$~3, and has evolved mainly passively until 
$z = 1.1$. 

In the (U-B)--rest frame (using BC03 stellar population models with solar metallicity and age equal to 4~Gyr), a scatter in $(i_{775}-z_{850})$ of $0.042\pm
0.010$ corresponds to $0.050 +/-  0.011$. As pointed out in van Dokkum (2000) and Blakeslee et al. (2003a), CMR scatters vary little with redshift. The Blakeslee et al. (2003a) scatter for the elliptical CMR in RDCS~1252--2927 ($0.024 \pm 0.008$), correspond to a scatter of $0.042 \pm 0.014$ in the (U-B)--rest frame, indistinguishable within the uncertainties from our result.

The scatter in the CMR for galaxies classified as S0
($\sigma_{int}=0.044\pm 0.020$) is
comparable to the one in the E CMR, but the galaxy colors are bluer
and are not compatible with a population that is as old as the
ellipticals.  All the S0s lie below the elliptical color--magnitude
relation.  In fact, between the elliptical and the S0 CMR fits there
is a zero point difference of $0.07 \pm 0.02$~mag, with the S0s being
bluer than the ellipticals.  One of the three S0/a galaxies has a color
that is 0.07~mag redder than the elliptical CMR, and another has a
color that is $\approx$~0.15 bluer than the CMR relation for Es. 
The inclusion of the S0/a galaxies does not significantly change the
fitted CMR for the S0s.

When we consider all galaxies within 2$\arcmin$ from the cluster center,
the S0 and total early--type slopes are similar, while the color offset
in the CMR is still present ($0.05 \pm 0.02$~mag). 
The S0 population of this cluster has a
very peculiar CMR with respect to the average cluster of galaxies.
In fact, for several other studies the CMR of the S0 population has a similar 
zero point and on average a larger scatter with respect to the
elliptical population (van Dokkum et al. 1998; Blakeslee et al. 2003a;
Holden et al. 2004; De Lucia et al. 2004), quite different from our results.
We will discuss this peculiar behavior in detail in the rest of the paper,
including an examination of orientation effects on the classification

In Fig.~\ref{ircolors}, the near-IR (Vega magnitudes) and $(i_{775} -
z_{850})$ (AB magnitudes) colors are shown compared with single burst stellar
population model predictions from BC03.  The S0 colors are consistent
with young ($<$~2~Gyr) solar metallicity, or older ($<$~3.5~Gyr), half
solar metallicity populations.  If the difference in E and S0 mean
colors is mainly due to metallicity, then even if the
two populations were formed at the same epoch, ellipticals must have been able
to retain more metals than the S0s, i.e., they were more massive at a
given luminosity (given the observed mass--metallicity relation for 
early-type galaxies, e.g., Tremonti et al. 2004, Bernardi et al. 2005, 
and references therein).  This would imply higher mass--to--light ratios for
the ellipticals with respect to the S0s.  However, the lack of
strong evolution in the slope and scatter of the CMR from the present out
to $z{\sim}1$ suggests that the CMR is mainly the result of a metallicity--mass
(i.e. metallicity--magnitude) relation 
(e.g., Kodama \& Arimoto 1997; Kauffman \& Charlot 1998
Vazdekis et al. 2001; Bernardi et al. 2005).
So, at a given magnitude we do
not expect large metallicity variations.

If the offset is due to a 
different star formation history, a model with solar metallicity
and with an exponential decay of the star formation will reproduce the offset
at a galaxy mean age of $\approx$~3.5~Gyr. This age is consistent with
the small scatter observed in the S0 CMR.
We would then be observing galaxies
that followed different star formation: single burst and passive 
evolution for the ellipticals and exponentially decaying star formation
for the S0s. An exponential decay in the star formation is observed 
in field spiral samples (Rowan--Robinson 2001). If this is the case, 
our S0 population might
be the evolved product of an old spiral population that was already in place
in this cluster when the ellipticals formed and then gradually lost
available gas for star formation.

If the E vs S0 color difference is mainly due to a difference in age,
for a solar metallicity and a single burst BC03 template with age 4~Gyr, 
the color difference
corresponds to an age difference of $\sim$1~Gyr.  For clusters of
galaxies at z~$>$~1, the cluster members on the red sequence are only a
part of all the progenitors of present--day early--type galaxies.
Some of today's galaxy progenitors would have been bluer than the red
sequence at these redshifts (van Dokkum \& Franx 2001). In the S0
population of this cluster, we may be seeing the transitional
progenitor population that in $\sim$1~Gyr will evolve onto the same
red sequence as now occupied by the ellipticals.

Either of the latter two scenarios would be consistent with 
the Postman et al. (2005) observed deficit
of the S0 population of our ACS cluster sample,
 when compared to lower 
redshift samples, implying that part of the S0 population is still forming in clusters at redshifts around unity.

\section{Galaxy shape properties}


Since the galaxies classified as S0 in RXJ~0910+5422 are found to be
systematically bluer (with respect to the red sequence) than the
S0 populations observed in previous studies, we wish to examine further the 
properties of these galaxies in terms of their shapes and light distributions,
and how they compare to the elliptical and spiral samples in this and other clusters.
The shape parameters that we will consider are Concentration and Asymmetry
(Abraham et al. 1996; Conselice et al. 2004), Sersic index $n$,
and galaxy axial ratios.

\subsection{Asymmetry and Concentration}

In Fig~\ref{cas} (left), we compare the Asymmetry A and the Concentration C
for ellipticals, S0s, and spirals with $(i_{775} - z_{850})$ colors between 0.5
and 1.2~mag. The Asymmetry parameter is obtained by subtracting a 180--degree
rotated image from each original galaxy image, summing the residuals and
including a correction for the background. The Concentration parameter is
defined as in Abraham et al. (1996) as the sum of the galaxy flux within an
aperture $r_{0.3}$ divided the total flux. $r_{0.3}$ is calculated using the
SExtractor fit to the galaxies at 1.5~$\sigma$ above the background. The
obtained semi--major and semi--minor axes from this fit were multiplied by
0.3 to derive the $r_{0.3}$ aperture (see also Homeier et al. 2005b).
Early--type and late--type galaxies lie on different regions in this A vs C
plane. All our red sequence S0s have $A < 0.3$ and $C > 0.3$. All but one
have $A < 0.2$ and $C > 0.3$.  This is the same locus in the A--C plane that
is occupied by most early--type galaxies in Abraham et al. (1996) and in our
Postman et al. (2005) low--redshift sample (Fig~\ref{cas}, right). This last
sample includes 5 strong lensing clusters observed as part of our ACS GTO
program [Zw1455+2232 ($z{=}0.258$), MS1008-1224 ($z{=}0.301$), MS1358+6245,
CL0016+1654 ($z{=}0.54$), and MS J0454-0300 ($z{=}0.55$)]. Visual and automated
classification for this cluster in the $i_{775}$--band for all galaxies with
$i_{775}< 22.5$ was performed by Postman et al. (2005).  We conclude that the
S0 population presents statistical parameters typical of an early--type
population (low asymmetry and high compactness).

\subsection{Sersic Indices}

Fig~\ref{justsersic} plots Sersic index $n$ as a function of galaxy effective
radius $R_e$, both from our GALFIT modeling, for
ellipticals, S0s and spirals with $(i_{775} - z_{850})$ colors between 0.5 and
1.3~mag.  Red sequence ($(i_{775} - z_{850})$ color between 0.8 and 1.1~mag)
galaxies are shown by large diamonds.  Most spirals have $n<2$ and
most early--types have $n>2$, as expected.  However, the $n$ values do not
permit us to discriminate between S0s and ellipticals in a unique way, unless they are combined with goodness--of--fit information for the Sersic model (e.g. the {\it Bumpiness} parameter introduced by Blakeslee et al. 2005)

\section{Axial Ratios}
\subsection{Axial Ratio Distribution}

In Fig~\ref{ab} we compare the apparent axial ratio from SExtractor
versus effective radii for
elliptical and S0s with $(i_{775} - z_{850})$ colors between 0.5 and 1.3~mag.
The axial ratios have been verified by using ELLIPROF (the isophotal
fitting software that is used for Surface Brightness Fluctuations analysis in
Tonry et al. 1997 and Mei et al. 2005) on each galaxy image, after the cleaning
procedure. 
As above, red sequence galaxies are shown by large diamonds.  The red sequence
ellipticals and the bluer S0s have different axial ratio distributions, 
with all
red sequence S0s showing axial ratios $\frac{b}{a} \lesssim 0.7$, and nearly
all red sequence ellipticals with $\frac{b}{a} > 0.7$.
  Assuming axisymmetric
disks (oblate ellipsoids) viewed with random orientation, and with a Gaussian distributed intrinsic axial ratio (with mean equal to $0.3 \pm 0.1$ (extreme thin disk),  $0.5 \pm 0.1$ (early--type galaxy) and  $0.75 \pm 0.1$ (elliptical) (Jorgensen \& Franx 1994)), one would expect $\gtrsim\,$40\%, $\gtrsim\,$60\%, and $\gtrsim\,$90\%, respectively, of the S0s to have axial ratios above 0.7. 
Just 9\% (1 out of 11)
of the red sequence S0s are observed to have an axial ratio this large 
(or 22\% for the full S0 sample in this cluster field) (Fig~\ref{ab2}; top).  
The random probability that the S0 axial ratios
would show such a low fraction with $\frac{b}{a} > 0.7$ is less than 1\%. 
This is a very simple model, but it points out a lack 
of round S0s,
indicating either that there is some orientation bias in the classification,
or that this class of objects is intrinsically prolate in shape.
Jorgensen \& Franx (1994) found a similar deficit of round S0s in the center of the Coma cluster. They concluded that part of the face--on S0s were classified as elliptical galaxies. Fabricant et al. (2000) also found a deficit of round S0s in the cluster CL~1358+62, at z=0.33. Their analysis shows that ellipticities and bulge--to--total--light--ratio do not allow us to distinguish elliptical from S0 galaxies.
The other two ACS GTO clusters at $z{\,>\,}1$
(RXJ1252.9-292 and RX~J0848+4452) do not show a similar lack of round S0s,
as 90\% of the red sequence ellipticals (out of $\approx\,$70) and 47\% of
the S0s (out of $\approx\,$35) have $\frac{b}{a} > 0.7$ (Fig~\ref{ab2}; bottom). 
 This bias is also not observed in other clusters of our ACS Intermediate Redshift Cluster Survey (see Fig~5 from Postman et al. 2005), for which  more than 40\% of the S0 galaxies have axial ratios above 0.7.

The observed peculiarity of the RXJ~0910+5422 
S0 axial ratio distribution might
call into question our result above that the S0s have a significant color
offset with respect to the ellipticals.  For instance, if there is a 
bias in our color measurement procedure which causes elongated objects
to have colors that are too blue, then the color offset found above may
be artificial.  Such a color bias might occur if the high inclination angles
bias our $R_e$ measurements to higher values, and if the S0s become
progressively bluer at larger radii.  We first examine this possibility,
then proceed to discuss resolutions to the peculiarity of the S0
axial ratio distribution, with the aim to establish if a misclassification of 
 face--on S0s as ellipticals would bias the measurement of the offset between the elliptical and S0 CMR zero points.


\subsubsection{Internal color gradients}

It is conceivable that our
$(i_{775} - z_{850})$ colors could be biased by aperture effects in
the nearly edge--on S0s, for which the (possibly) bluer outer disks might
contribute more to the galaxy colors than in the rounder ellipticals population.
If this effect were severe enough, it might mimic the offset in color of the
S0s and ellipticals found above.  We test this possibility here.
S0 and elliptical internal color profiles are
shown in Fig~\ref{gradso} and Fig~\ref{grade}.  The gradients have
been calculated with aperture photometry on the same images used to calculate
our $(i_{775} - z_{850})$ colors.  Circles are $(i_{775} - z_{850})$ colors at
different radii, the cross the $(i_{775} - z_{850})$ color at the effective
radius, used for the CMR.

The S0 colors profiles do not show strong gradients. In particular,
the $(i_{775} - z_{850})$ colors calculated at the effective radii are
not systematically bluer than colors determined at smaller radii. 
Most of S0 galaxies have flat profiles; one has a blue inward gradient
(ACS ID 1621; $z_{850}=23.72$~mag, $(i_{775} - z_{850})=1$~mag). 
In two galaxies (ACS ID 1393 and 3177) 
the colors in the central 0.15~$\arcsec^2$ are redder than the color
at the effective radius.  When compared with elliptical gradients, on average S0
colors do not appear biased towards higher effective radii and bluer colors than
the ellipticals.  
We note that three elliptical galaxies show blue inward gradients (ACS ID
1753, 1519, 3323).

\subsubsection{Orientation or intrinsic shape: Axial ratios vs $(i_{775} - z_{850})$ colors}

Orientation biases are known to occur in the classification
of ellipticals and S0s in local galaxy samples.  For instance, 
Rix \& White (1990, 1992) showed, based on both isophotal and
dynamical modeling, that a large fraction of ellipticals contain
a disk component with at least $\sim\,$20\% of the light, 
but which is hidden due to projection effects.
Jorgensen \& Franx (1994) found a strong deficit of round S0s
in a sample of 171 galaxies in the central square degree of the
nearby Coma cluster, and concluded that inclination angle played 
a large part in the classification of Es and S0s.  Michard (1994)
proposed that, except for the bright boxy ellipticals without 
rotational support, early-type galaxies comprise a single class
of oblate rotators with orientation being the main criterion for
classification as either E or S0.  On the other hand, van den Bergh (1994)
explained the predominance of flattened S0s by invoking two distinct
subpopulations: bright disky objects intermediate between ellipticals 
and spirals, and a fainter population of prolate objects.

We now address the question of whether the galaxies classified as
S0 in RXJ~0910+5422 are preferentially flattened in shape because
of an orientation bias in the classifications or intrinsically prolate
shapes.  If it is an orientation bias,
then this could mean either that (1) face-on S0s have been misclassified as
ellipticals because their disks are not apparent, or (2) that edge-on spirals
tend to be called S0s because the spiral structure is obscured.
Either would result in a predominance of flattened S0s.
However, in the former case, the misclassification
of face-on S0s as ellipticals would tend to blur any color separation
between the two classes, while in the latter case, a color offset
might be introduced between the two classes because of contamination
by bluer spirals.  Because we do observe a color offset between galaxies
classified as E and S0, with the S0s being bluer, 
it is possible to look for the ``missing'' population of face-on blue
galaxies by examining ellipticity versus color.   If a population 
of round blue objects is found, we can then determine the nature of
the classification bias, and whether it biases our color offset measurement.

Histograms of the axial ratio distributions for ellipticals and S0s 
are shown in the top panel of Fig.~\ref{ab2}.  
We find that 60\% of the early-type red sequence galaxies have
$\frac{b}{a} > 0.7$, but  95\% of these low-ellipticity galaxies
are classified as Es.
However, if we split galaxies instead by color, 
using $(i_{775} - z_{850}) < 0.99$~mag as the separation point,
then we find that 
54\% of all galaxies bluer than this separation have $\frac{b}{a} > 0.7$, 
while 43\% of the red sequence galaxies bluer than this 
[$0.8 < (i_{775} - z_{850}) < 0.99$~mag] have $\frac{b}{a} > 0.7$
(Fig.~\ref{ab3}).  
Thus, the deficit of round S0 galaxies (which are also significantly bluer
than the mean of the E class) is not found when the
early-type galaxies are split based purely on color.  This suggests that
some of the bluer round galaxies are the face-on counterparts of those
classified as S0.

Fig~\ref{blue} shows axial ratios vs $(i_{775}- z_{850})$ color residuals
with respect to the total early-type galaxy CMR relation, for all galaxies
in the RXJ~0910+5422 red sequence.  Galaxy types are coded with different
symbols; using the color residuals in this way takes out the effect of the
magnitude dependence of the colors.  There are five round blue ellipticals,
i.e., with $\frac{b}{a} > 0.7$ and on the blue side of the early-type CMR.
If these are the face-on counterparts of the blue S0s, then the S0
axial ratio distribution becomes much more in line with expectations
for a randomly oriented disk population (35\% are rounder than 
$\frac{b}{a}{\,=\,}0.7$).  Further, we note that
four other ellipticals were classified E/S0 in Postman et al.\ (2005);
if these are also taken as face-on S0s, then the axial ratio distribution
comes in very close agreement with expectations.
We conclude that the face-on S0s in RXJ~0910+5422 are classified as
ellipticals, just as in local early-type galaxy samples.

From our axial ratio simulations and statistics from our z$\approx$1 
ACS GT0 sample, the number of blue S0
galaxies might be between 25 and 40\% higher than estimated in 
section~5.3.1. Moreover, the inclusion of a few blue S0s in the elliptical
red--sequence will have the effect of increasing very slightly the observed
CMR scatter for the ellipticals.
The uncertainty in the Postman et al. (2005) S0  fraction in RXJ~0910+5422 is significantly larger than 40\% and hence a  systematic change by this amount for one cluster does not alter any of  the results or conclusions in that paper.

Throughout the paper,
we continue to use elliptical and S0 classifications from Postman et al. (2005), but keeping in mind the
presence of this possible projection effect. 

\section{Color trends, velocity dispersion and merging activity}

Stanford et al. (2002) have analyzed the X--ray and near-IR
properties of this cluster. This system appears to be fairly
relaxed based on its regular X--ray profile; however, they find
 indications that the cluster is in an early phase of formation. In fact, the
Chandra ACIS data show evidence for temperature structure, possibly 
due to an infalling group or mass streaming along a filament. 
The soft component of the X--ray emission (0.5--2~keV) dominates the
X--ray center of the cluster, while to the south there is a harder
component (2--6~keV) (see Fig. 6 from Stanford et al. 2002).
The cluster does not have a central BCG or cD galaxy, and the X--ray
emission center does not correspond to an optical grouping of galaxies;
rather a number of luminous confirmed cluster members are linearly
distributed, at least as projected on the sky (as shown in
Fig~\ref{cluster}).  We do not observe any strong trend of the galaxy
color $(i_{775} - z_{850})$ with distance from the cluster X--ray center
(Fig.~\ref{colors}).  Surprisingly, though not statistically
significant, the bluer galaxies are concentrated toward the cluster
center, instead of the outskirts, as in the other ACS intermediate
redshift cluster sample (Demarco et al. 2005; Goto et al. 2005; 
Homeier et al. 2005; 
Postman et al. 2005). This tendency might
support the hypothesis that we are observing sub-groups of galaxies in
an edge--on sheet, e.g a group of bluer disk galaxies on a redder, older population of ellipticals. 

There are 10 confirmed elliptical and 5 confirmed S0 members in the 
center of the cluster (R~$<$~500~kpc).  The average S0 redshift is
1.102~$\pm$~0.002 and the average elliptical redshift is
1.105~$\pm$~0.007 (the given uncertainties are standard deviations
around the mean). This indicates small relative velocities (the two
redshifts are indistinguishable given the errors) between the
classes and is true regardless of the possible classification
bias discussed above.  Unfortunately, our
spectroscopic sample does not permit us to track the cluster central
structure in detail.  The average relative velocity between the
confirmed E and S0s of $\approx$~500~km/s (that also corresponds to
the cluster velocity dispersion; see below) is fairly small.  If
merging of two distinct groups of galaxies is happening along the 
line of sight, we expect much higher
velocity dispersions and/or relative velocities between the
infalling S0s and the ellipticals.


Stanford et al. (2002) also suggest that active galaxy-galaxy merging
should be observed, based on the X--ray temperature structure.  
To investigate any on--going
dynamical activity, we calculated the cluster velocity dispersion, and
the merger rate.  From the 25 spectroscopically confirmed members (all galaxy types included) the
line--of--sight rest--frame velocity dispersion is $\sigma =
675 \pm 190$~km/s, using the software ROSTAT from 
Beers, Flynn, \& Gebhardt (1990).
The available Chandra data give X-ray temperatures ranging from $kT =
7.2^{+ 2.2}_{-1.4}$~keV (Stanford et al.\ 2002) to $kT = 6.6^{+
1.7}_{-1.3}$~keV (Ettori et al.\ 2004). Wu et al. (1999; see also
Rosati et al. 2002) gives the relationship between $kT$ and $\sigma$
for relaxed clusters, which predicts that the velocity dispersion 
corresponding to the measured X-ray temperature should be
$\approx$~1000~km/s, considerably higher than we have found.
Again, in the case of merging groups (along the line of sight) 
we would also have expected a higher velocity dispersion.

\section{Red galaxy pairs}

\begin{figure*}
\includegraphics[angle=-90,scale=0.7]{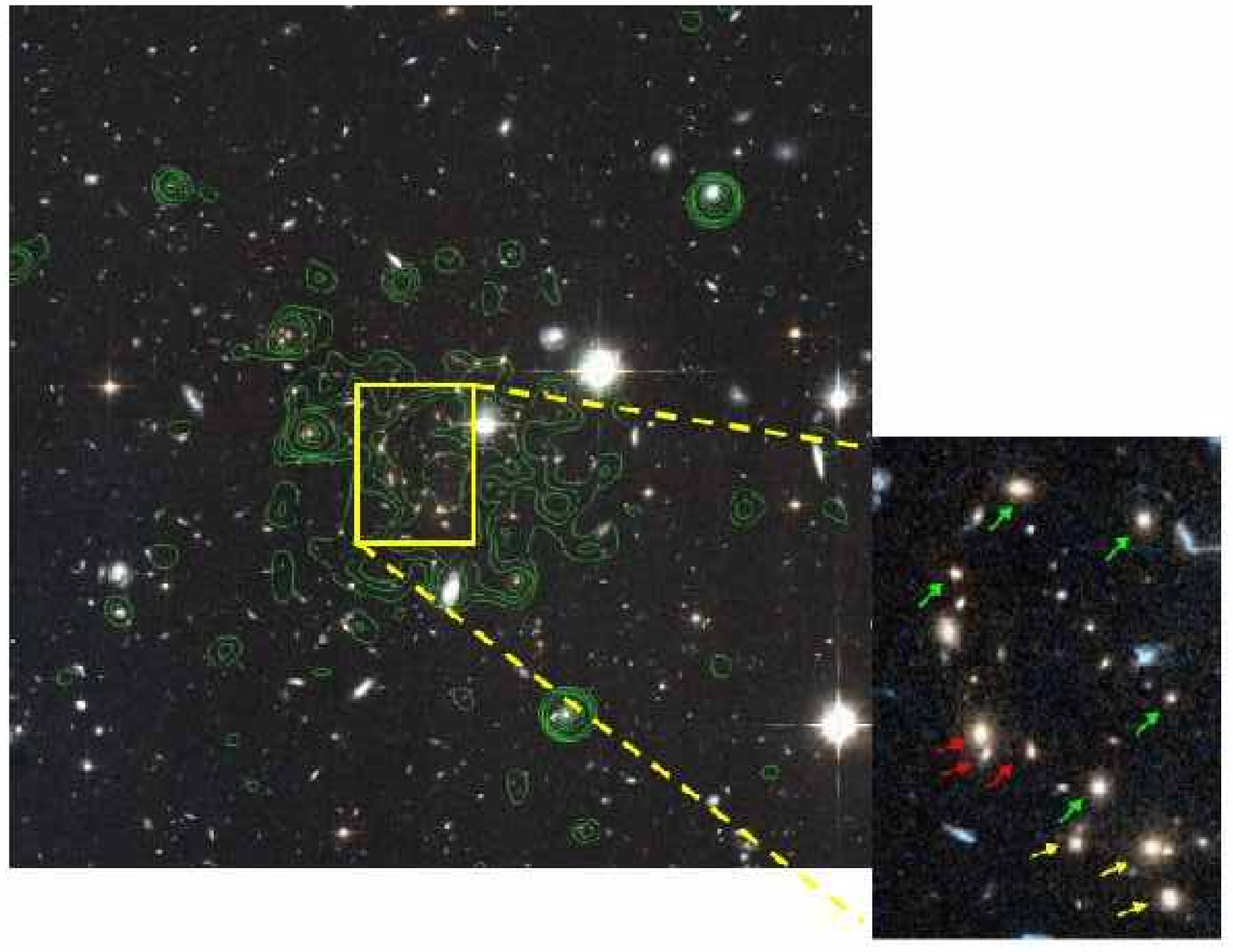}
\caption {The ACS color image with X--ray contours overlaid. The X--ray observations are from Chandra ACIS-I, over the energy range of 0.5--2KeV, and were smoothed with a 5~arcsec FWHM Gaussian. North is up, East on the left. In the enlargement, galaxy pairs in the
central filamentary structure are shown. 
Two main groups of interacting galaxies 
lie along the filamentary structure at the center of the cluster.  The
first group is shown by the yellow arrow, the second one by the red
arrows.  Three other red--sequence early--type galaxies are shown by the green
arrows. {\label{cluster}}}
\end{figure*}

\begin{figure*}
\includegraphics[scale=0.7]{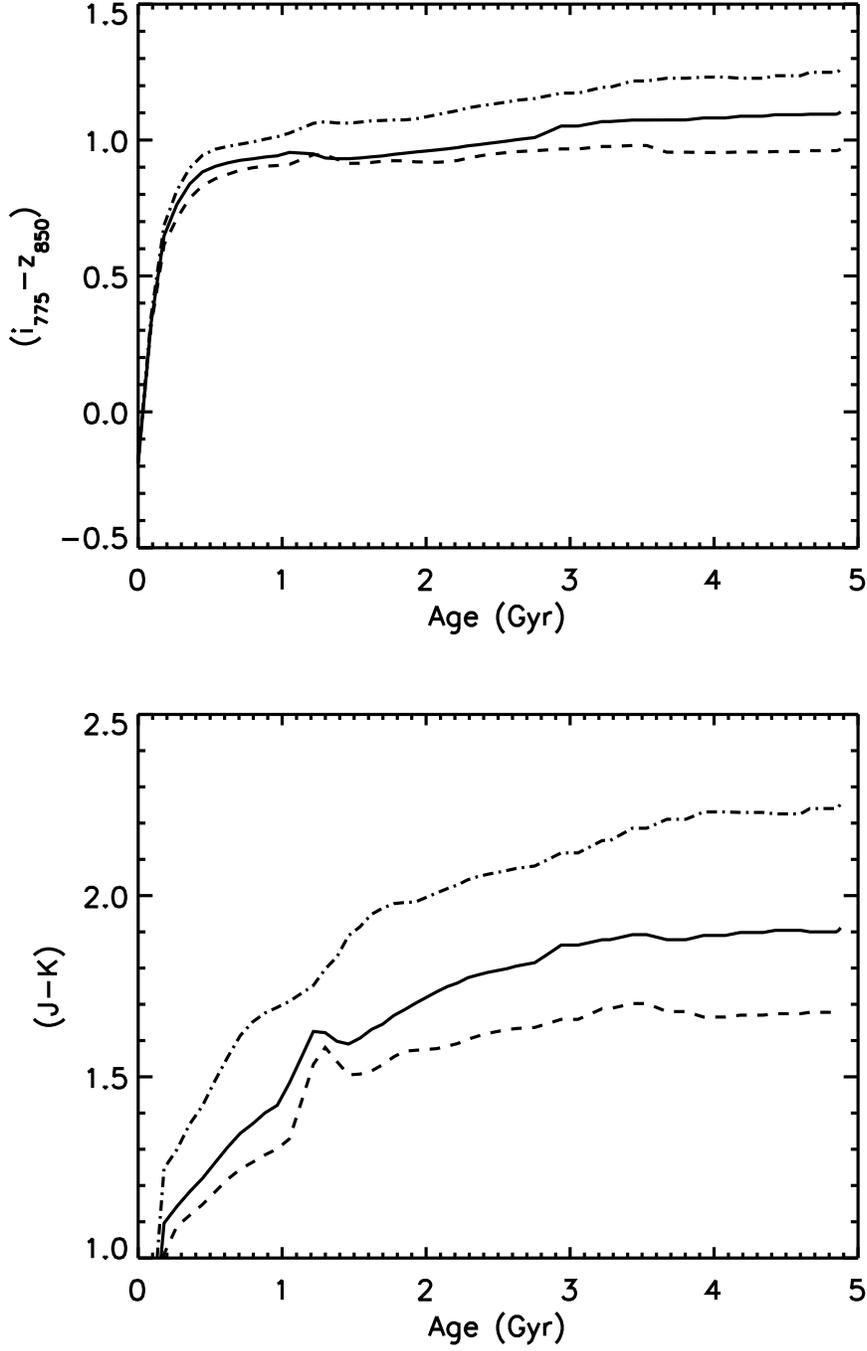}
\caption{ $(i_{775} - z_{850})$ and (J-K) colors as a function of galaxy
age. BC03 solar metallicity, age equal to 4~Gyr models are shown with a solid line. The dashed lines
are for half solar and the dot--dashed lines for twice solar metallicity.
We considered a sample of color--selected galaxies with $0.8<(i_{775} - z_{850})<1.1$~mag and $(J-K)>1.45$.
 {\label{colage}}}
\end{figure*}

\begin{figure*}
\includegraphics[angle=90,scale=0.6]{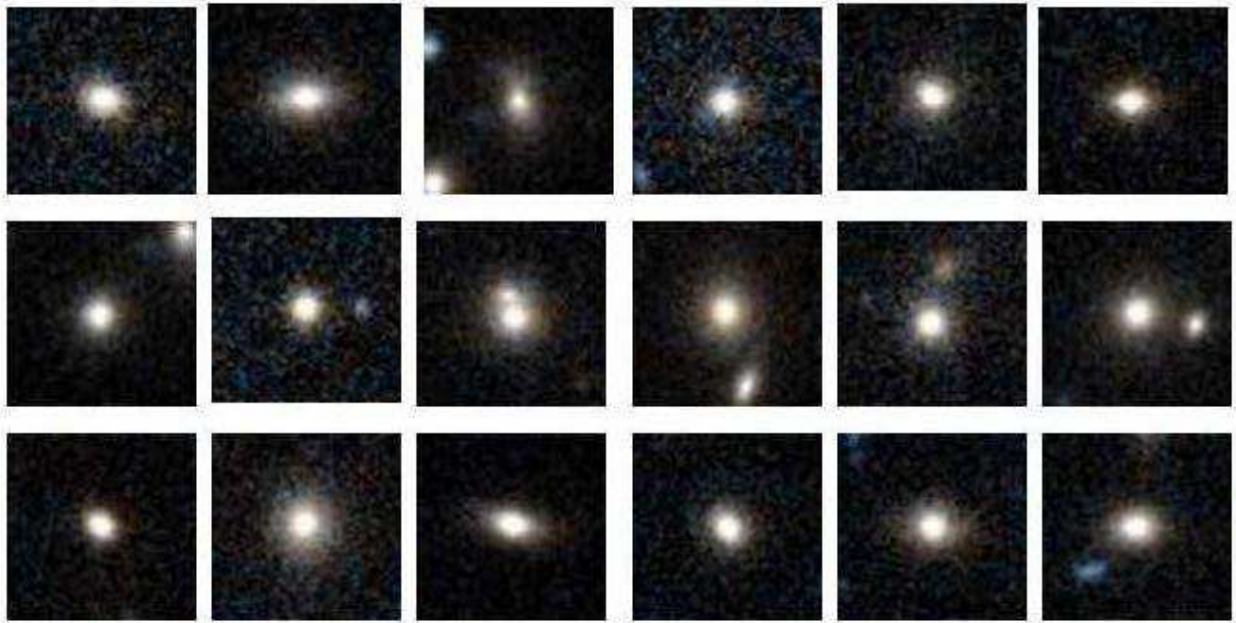}
\caption{ Color images of the 20 CMR ellipticals
within 2~$\arcmin$ from the cluster center. 
Two of the CMR faint ellipticals, that are satellites of brighter galaxies, are shown in the same postage image of their bright companion.
 {\label{epostage}}}
\end{figure*}

\begin{figure*}
\includegraphics[angle=-90,scale=0.6]{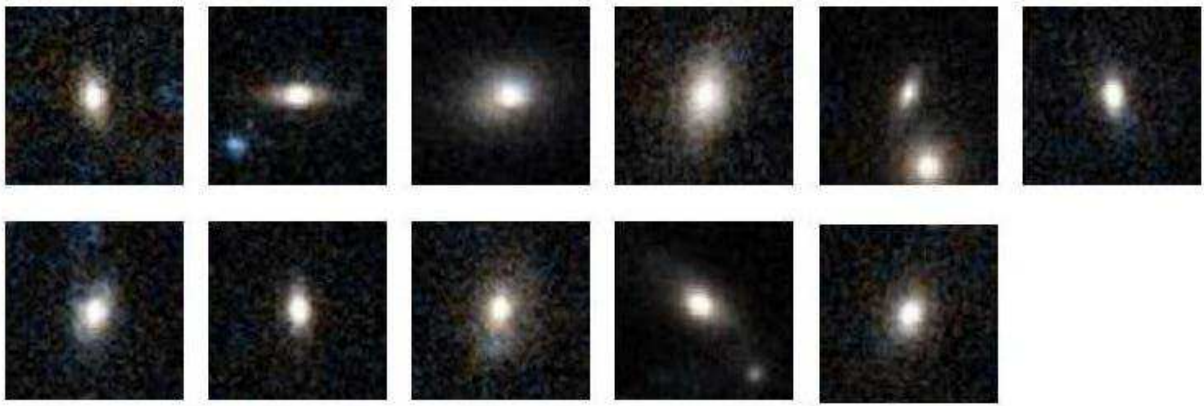}
\caption{ Color images of the 11 CMR S0 galaxies
within 2~$\arcmin$ from the cluster center.
 {\label{sopostage}}}
\end{figure*}

\begin{figure*}
\includegraphics[angle=90,scale=0.4]{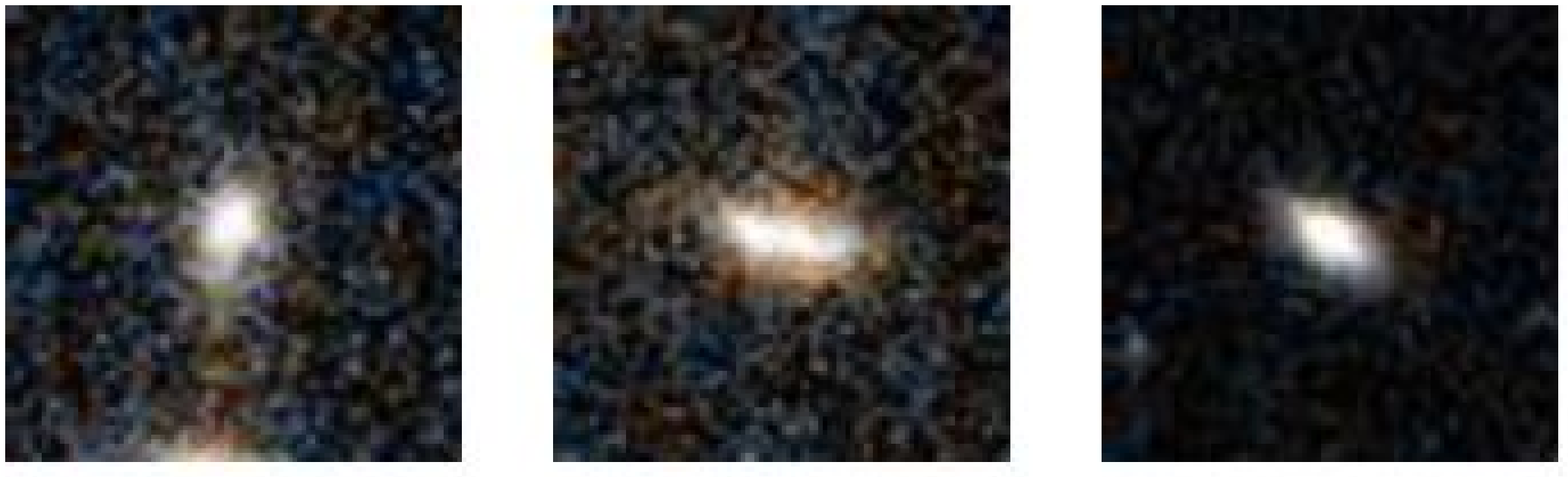}
\caption{ Color images of the three CMR S0/a galaxies
within 2~$\arcmin$ from the cluster center.
 {\label{sapostage}}}
\end{figure*}

\begin{figure*}
\includegraphics[angle=90,scale=0.4]{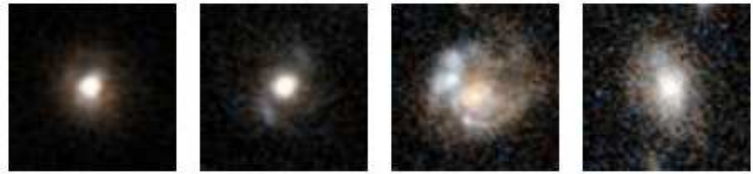}
\caption{ Color images of the four bright spirals
with $z_{850}$ brighter than 22.5~mag and $(i_{775}-z_{850})$
between 0.5 and 1.3~mag,
within 2~$\arcmin$ from the cluster center. 
These spirals have luminosities
that are similar to those of the red--sequence bright ellipticals.
 {\label{spipostage}}}
\end{figure*}

\begin{figure*}
\includegraphics[angle=90,scale=0.7]{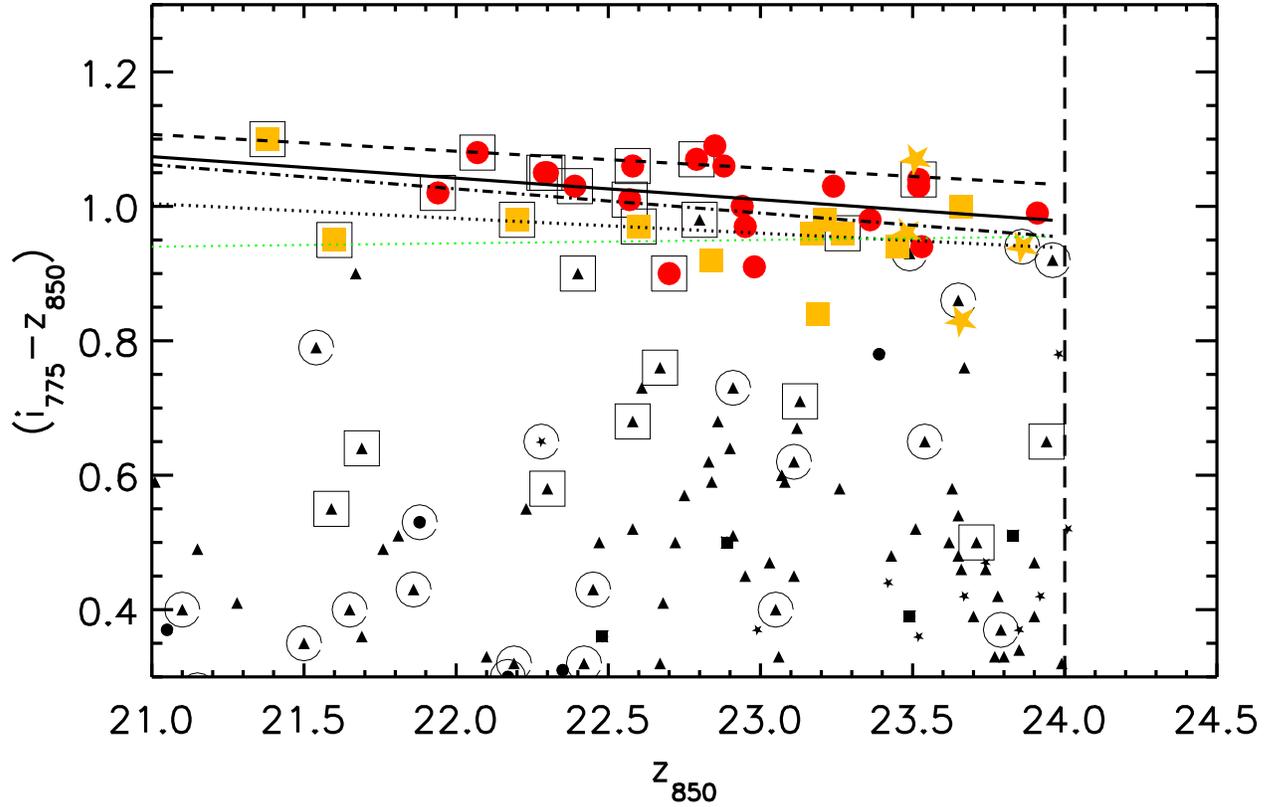}
\caption {The color-magnitude relations for RXJ~0910+5422. Red circles and yellow squares are respectively E and S0 candidates on the red sequence. Yellow stars are S0/a candidates on the sequence.
Smaller black circles, squares, and stars are used for the corresponding
early--type galaxies that do not
lie on the red sequence. Small triangles are late-type galaxies. Boxes
are plotted around confirmed cluster members. Open circles are
plotted around confirmed interlopers.  The solid line is the fit
to the ellipticals, the black dotted line the fit to the S0s, and the
dashed--dotted line the fit to the entire sample within 2$\arcmin$. 
 The green dotted is fitted to the S0s within 1$\arcmin$. The dashed line is
the RXJ1252.9-292 color--magnitude relation scaled to the redshift
and colors for RX0910+5422. The long--dashed vertical line is the magnitude limit of
the morphological classification $z_{850} = 24$~mag. 
The S0 CMR is bluer than the elliptical CMR by $0.07 \pm 0.02$~mag. {\label{cmd}}}
\end{figure*}

\begin{figure*}
\includegraphics[scale=0.7]{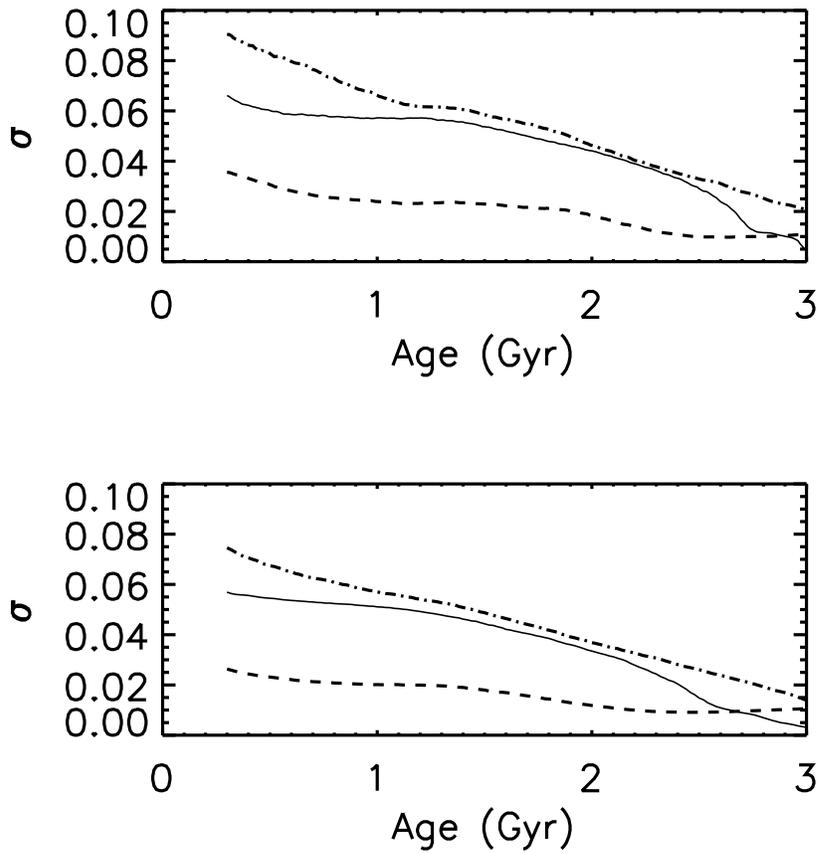}
\caption {CMR scatter as a function of galaxy age from
simulation obtained using BC03 single burst (on the top) 
and constant star formation (on the bottom) stellar population models
for solar metallicity (solid line), half solar (dashed line) and twice
solar (dash--dotted line). From the measured scatter it is possible
to estimate the mean age of the stellar bursts.  {\label{scatterage}}}
\end{figure*}

\begin{figure*}
\includegraphics[scale=0.7]{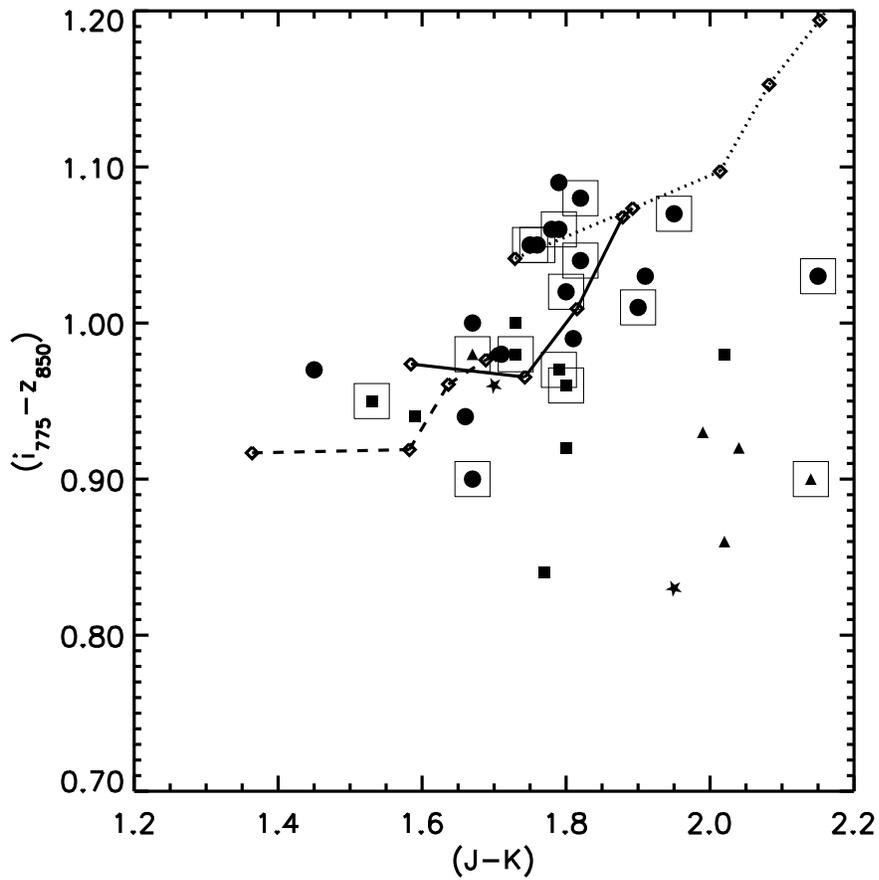}
\caption {$(i_{775} - z_{850})$ (AB) vs $(J-K)$ (Vega) color. 
Filled circles are ellipticals, squares are S0s, and stars S0/as. BC03
single burst models are shown for half solar (dashed line), solar
(solid line), and twice solar (dotted line) metallicities. Ages
run from 1.5 to 3.5~Gyr, by steps of 0.5~Gyr, as diamonds from left to
right. The bluer S0 colors might be due to younger ages or lower
metallicities. {\label{ircolors}}}
\end{figure*}

\begin{figure*}
\includegraphics[angle=90,scale=0.7]{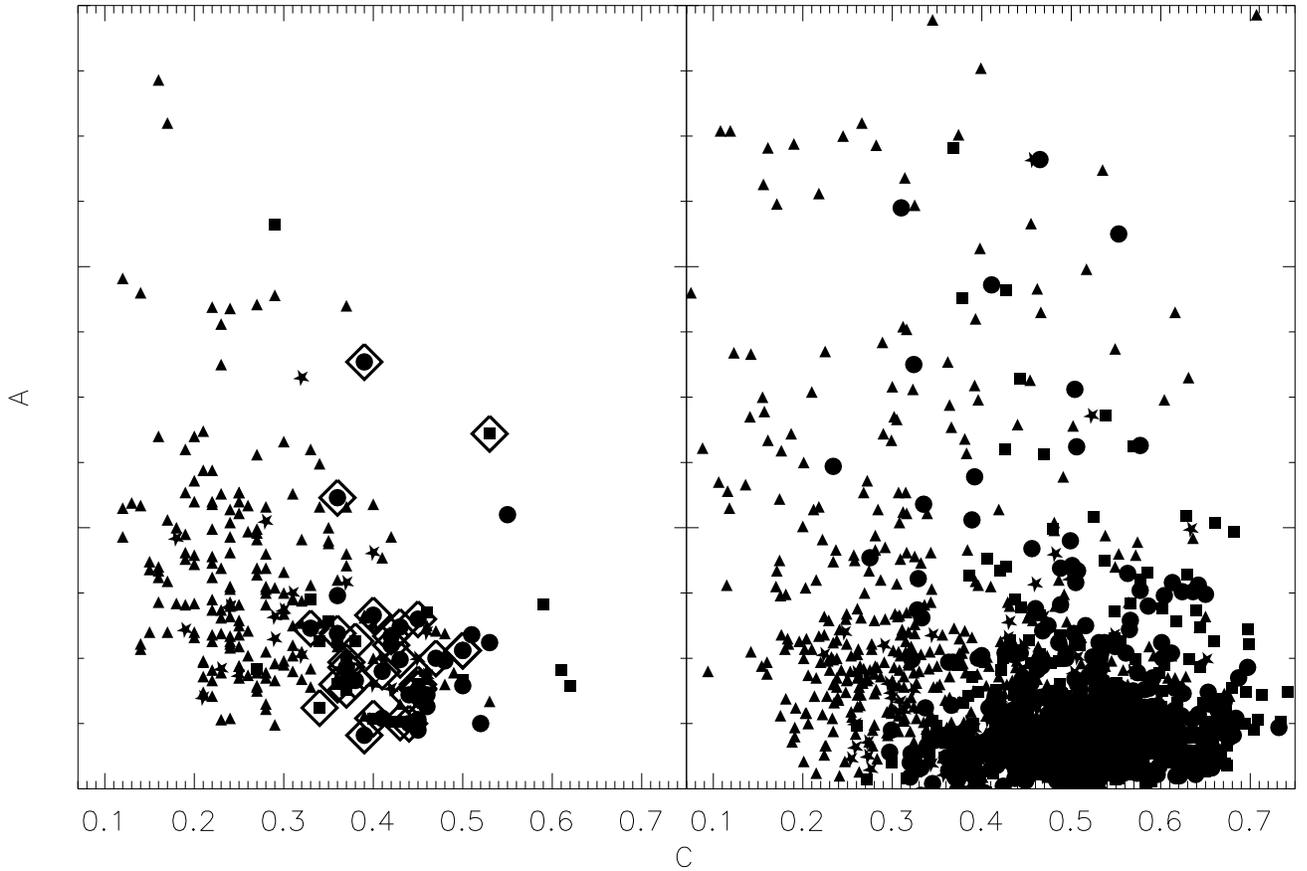}
\caption  { Asymmetry (A) vs Concentration (C) parameters for  ellipticals (circles), S0 (squares), S0/a (stars) and spirals (triangles) in the $(i_{775} - z_{850})$ color range between 0.5 and 1.3~mag for RXJ~0910+5422 (left). Large diamonds identify red sequence [$(i_{775} - z_{850})$ color between 0.8 and 1.1~mag] early--type galaxies. On the right, the same parameters are shown for all galaxies in our low--redshift sample using the same symbols for different galaxy types. E and S0 in RXJ~0910+5422 show parameters that are characteristic of an early--type population, A~$<$~0.2 and C~$>$0.3, as most of the low redshift early--type galaxies. This means that S0s were not likely misclassified late--type galaxies.  {\label{cas}}}\end{figure*}

\begin{figure*}
\includegraphics[scale=0.7]{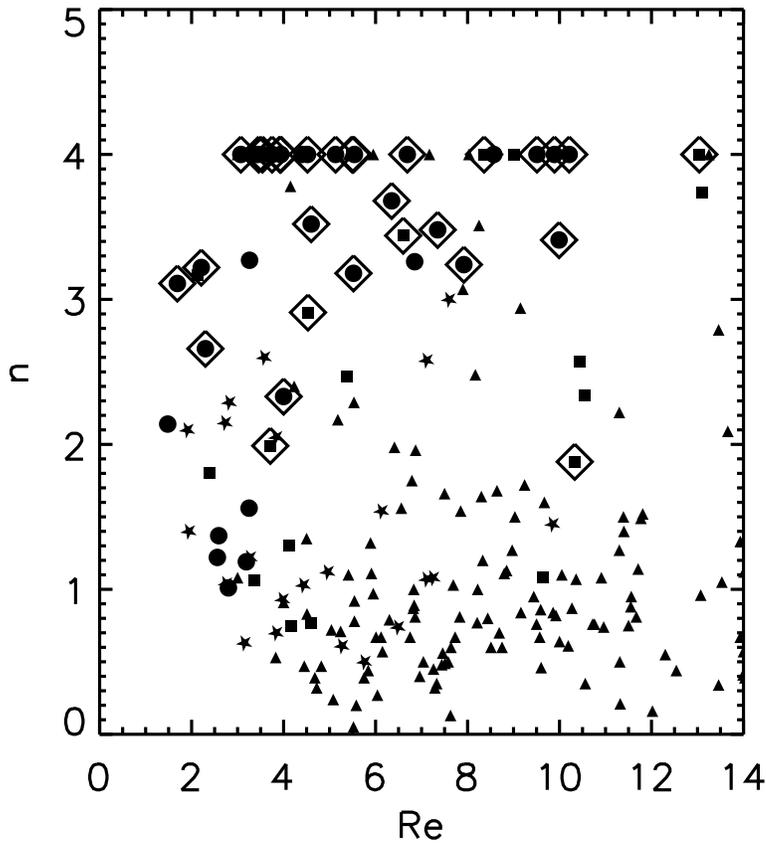}
\caption {Sersic index $n$ as a function of effective radius $R_e$
from GALFIT is shown for ellipticals (circles), S0s (squares), S0/a (stars) and spirals (triangles) with  $(i_{775} - z_{850})$ colors between 0.5 and 1.3~mag. 
Red sequence ($(i_{775} - z_{850})$ color between 0.8 and 1.1~mag) galaxies are shown by large diamonds. E and S0 in RXJ~0910+5422 show a Sersic index $\le$~2. This means that S0s were not likely misclassified late--type galaxies.  {\label{justsersic}}}
\end{figure*}

\begin{figure*}
\includegraphics[scale=0.7]{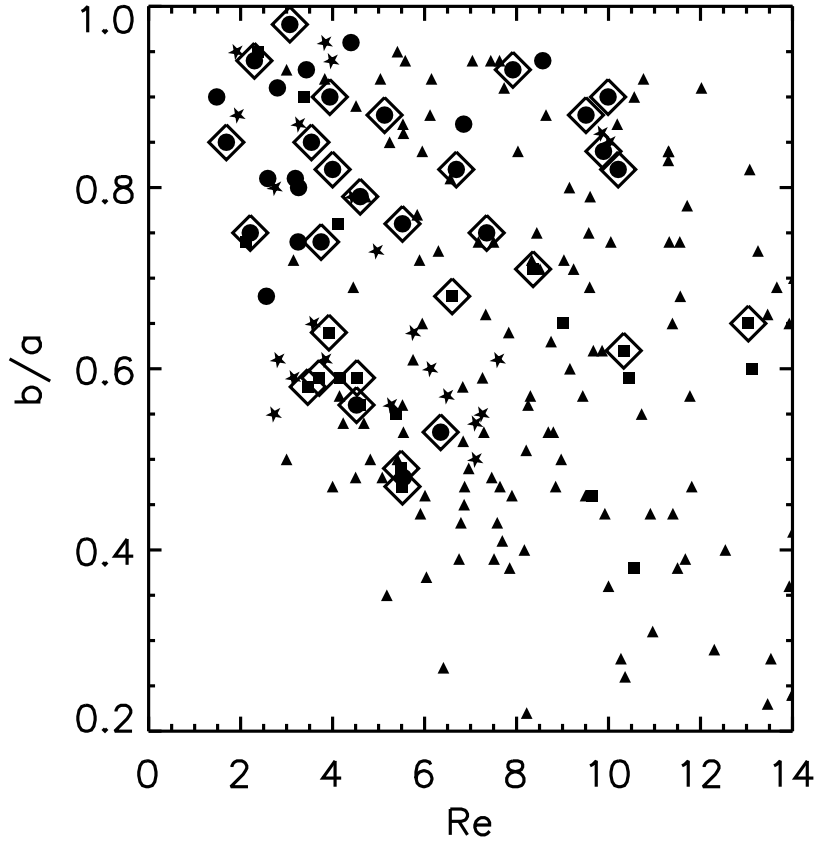}
\caption {Axial ratios vs effective radii are shown for ellipticals (circles), S0 (squares) in the ($(i_{775} - z_{850})$ color range between 0.5 and 1.3~mag). Large diamonds identify red sequence galaxies with $(i_{775} - z_{850})$ color between 0.8 and 1.1~mag. Stars are S0/as and triangles are spirals.   {\label{ab}}}
\end{figure*}

\begin{figure*}
\centerline{\includegraphics[scale=0.5]{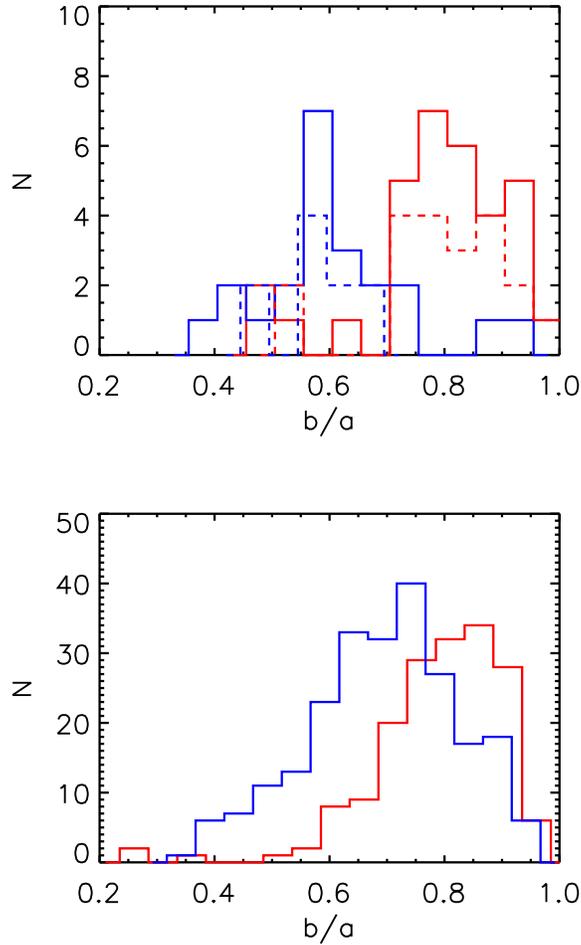}}
\caption {On the top, the histogram of axial ratios for the ellipticals (red line) and S0s (blue line) are shown. 
Dashed lines are for red sequence galaxies ($(i_{775} - z_{850})$ between 0.8 and 1.1~mag). 
 We notice a lack of high axial ratio (low ellipticity) S0s in RXJ\,0910+5422. On the bottom, we show elliptical (red) and S0s (blue) distributions in RXJ1252.9-292 and RX~J0848+4452. The median of their S0 distribution is consistent with that
expected for nearly axisymmetric disks viewed at random
inclination angles. {\label{ab2}}}
\end{figure*}

\begin{figure*}
\includegraphics[scale=0.7]{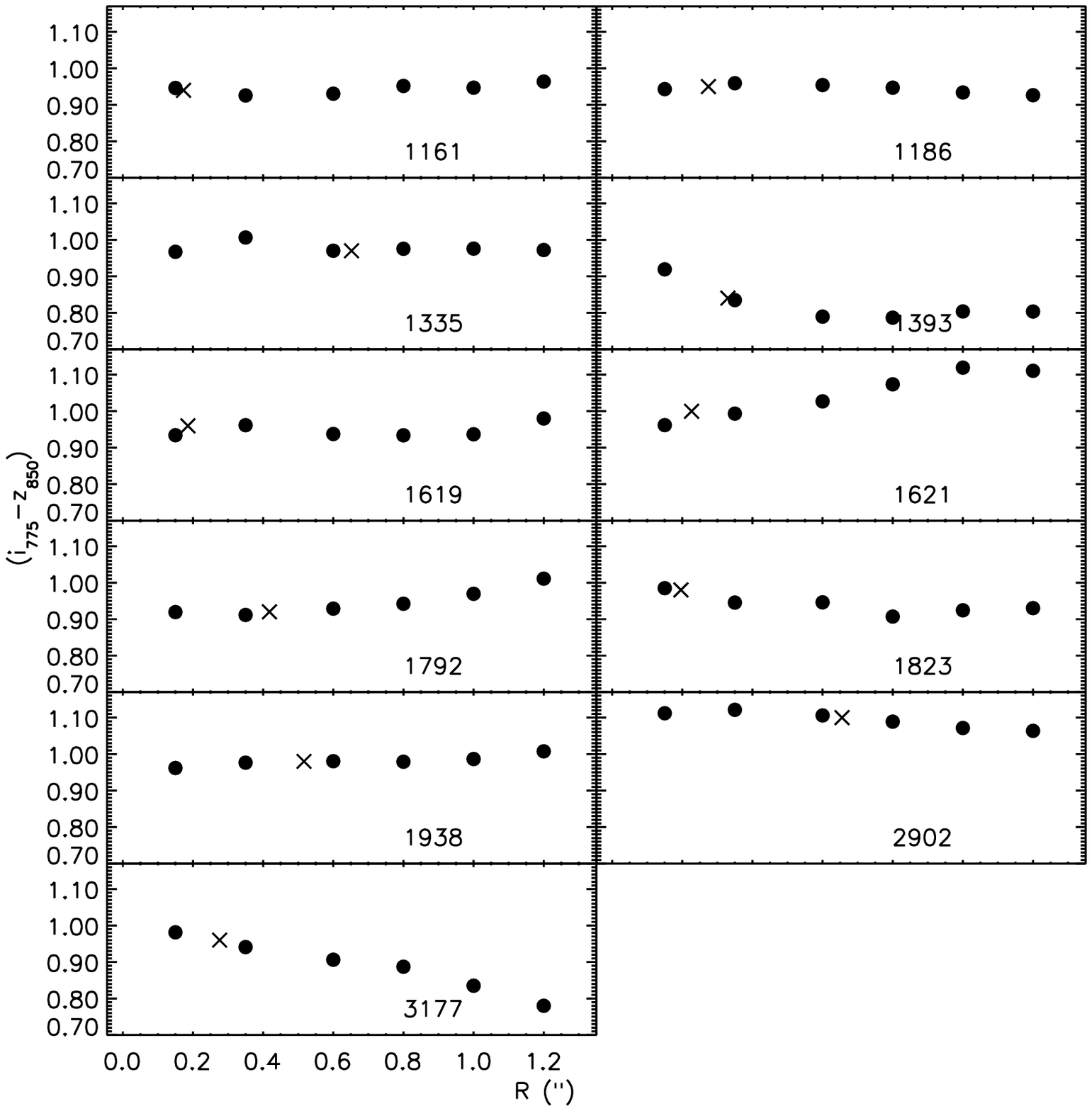}
\caption {Red sequence S0 color gradients as a function of distance from the galaxy center.
The cross is the color calculated at the effective radius. {\label{gradso}}}
\end{figure*}

\begin{figure*}
\includegraphics[scale=0.7]{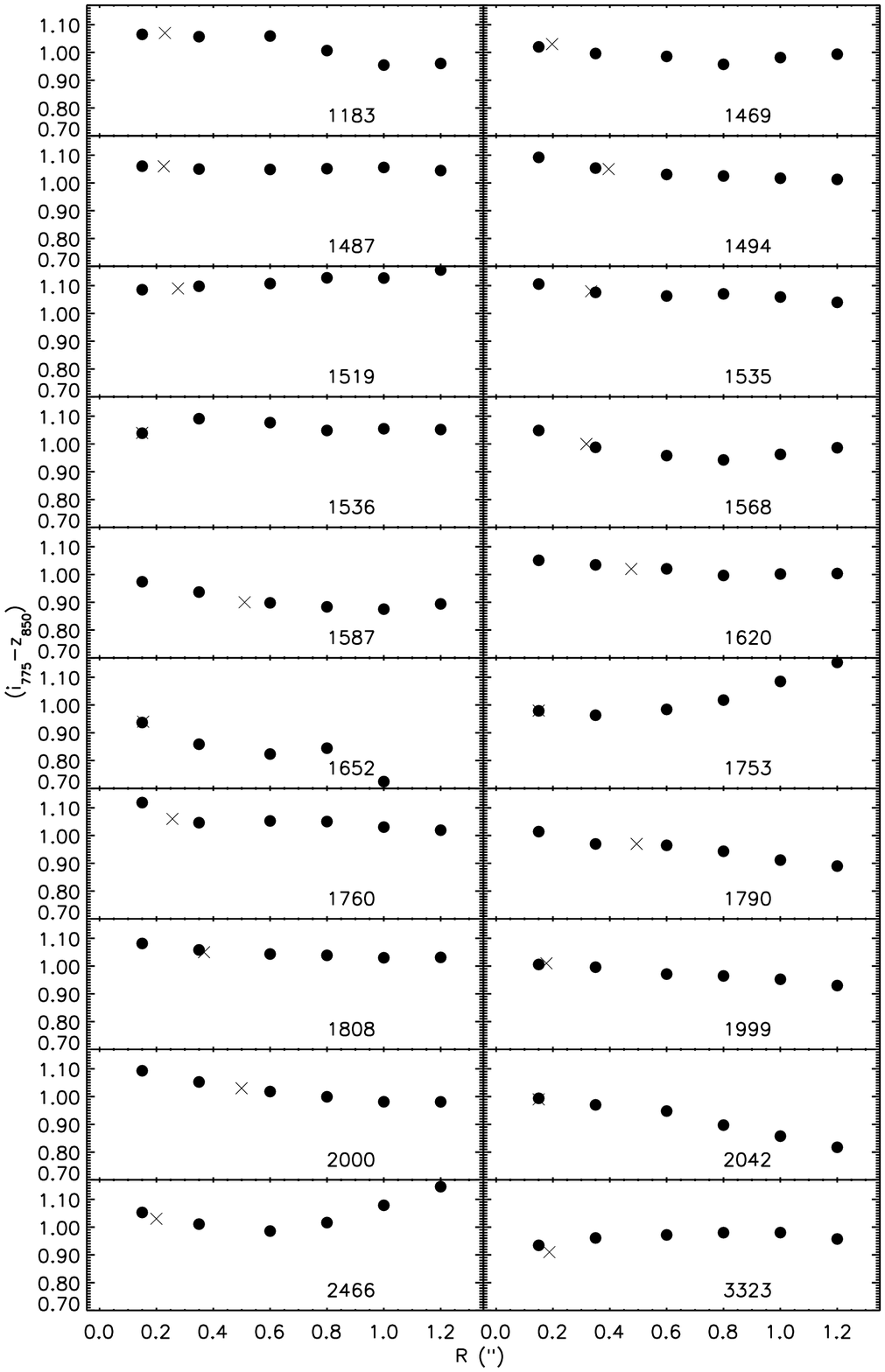}
\caption {Red sequence elliptical color gradients as a function of distance from the galaxy center.
The cross is the color calculated at the effective radius. {\label{grade}}}
\end{figure*}

\begin{figure*}
\centerline{\includegraphics[scale=0.7]{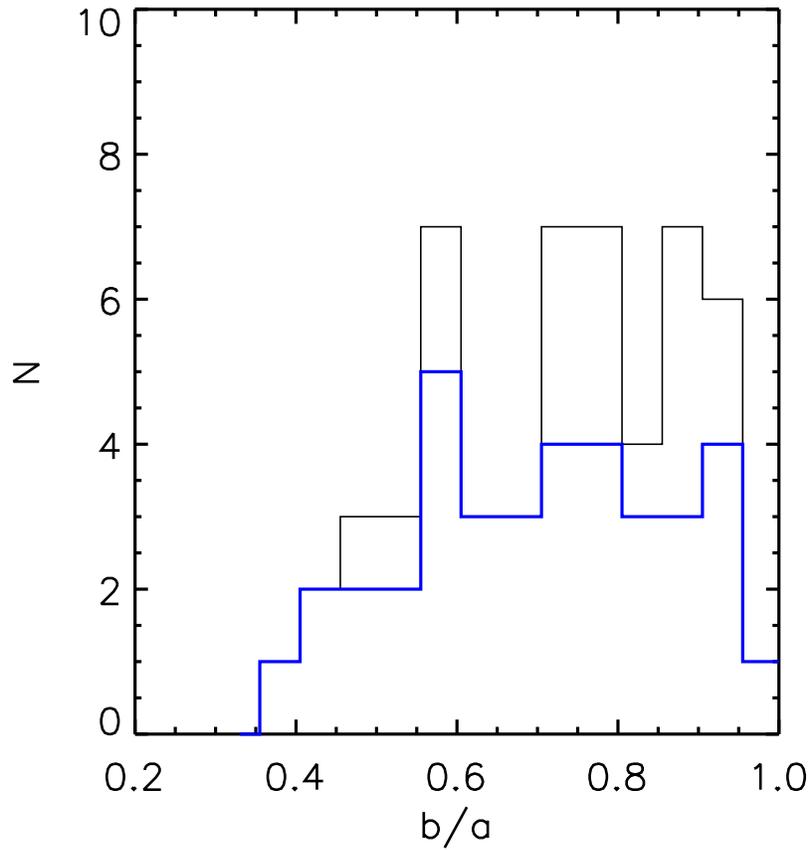}}
\caption {We show the histogram of the axial ratio of all early-types (black) and early-types with $(i_{775} - z_{850}) < 0.99$~mag (blue). Elliptical  
Whereas there is a lack of high axial ratio (low ellipticity) S0s in RXJ\,0910+5422, there is no similar lack in total early--type or in the blue early-type sample.  {\label{ab3}}}
\end{figure*}

\begin{figure*}
\includegraphics[scale=0.7]{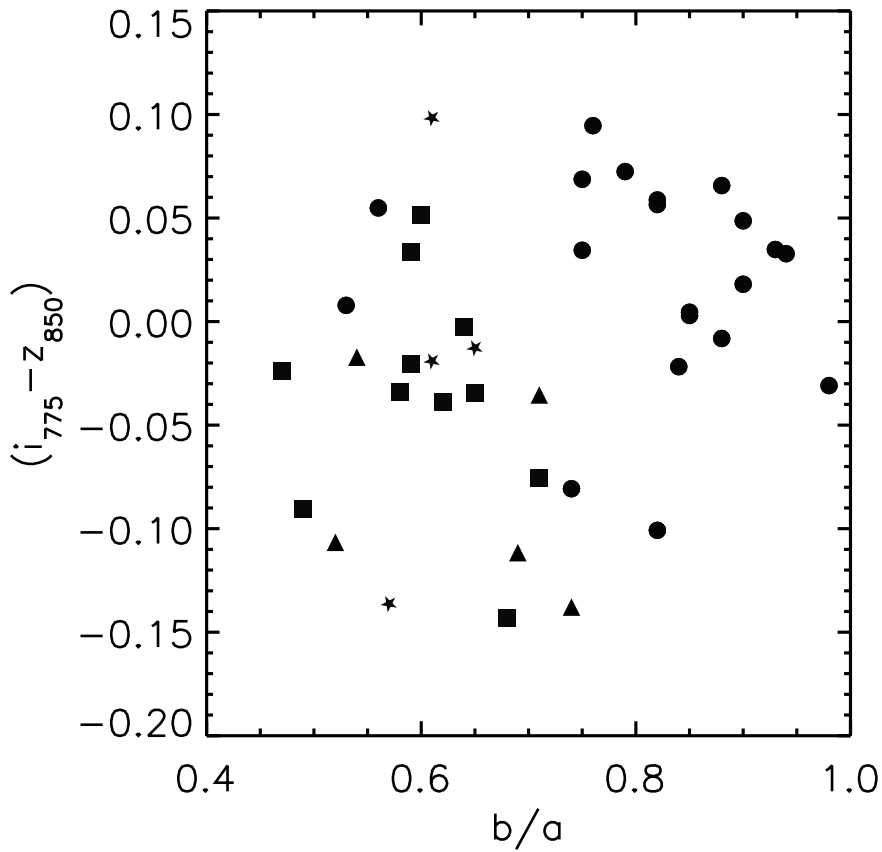}
\caption{ The $(i_{775} - z_{850})$ color residuals (with respect to the mean of full early-type CMR fit)
is plotted versus axial ratios for red sequence ($(i_{775} - z_{850})$ color between 0.8 and 1.1~mag) ellipticals (circles), S0s (squares), S0/a (stars) and spirals (triangles). 
While there is a lack of S0s with $\frac{b}{a} > 0.7$, we find five ellipticals and one spiral
that have blue colors (negative CMR residuals) similar to most S0s, but round
shapes ($\frac{b}{a} > 0.7$), unlike the classified S0s.  These may be the face-on counterparts
of the elongated blue S0 population (see text). {\label{blue}}}
\end{figure*}

\begin{figure*}
\includegraphics[scale=0.7]{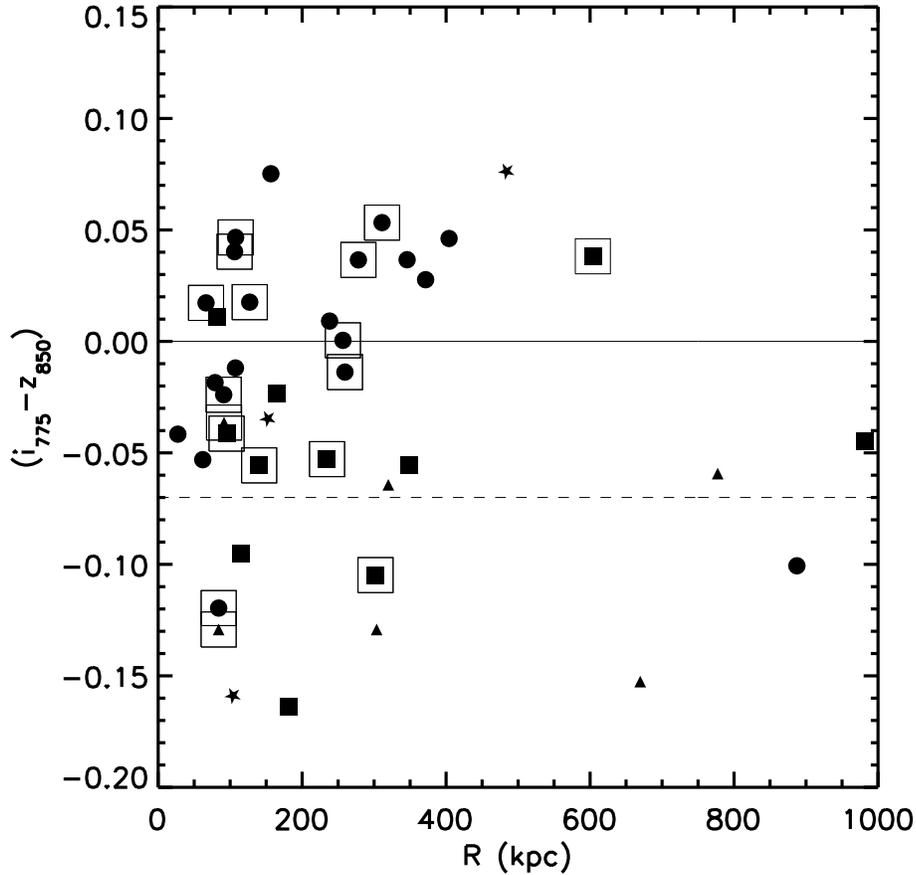}
\caption {Red sequence galaxy colors, corrected by the elliptical CMR, as a function of the distance from the cluster center. Filled circles are ellipticals, squares S0s, stars S0/a, and triangles spirals. Spectroscopically confirmed members are indicated by boxes.
Known non-members are omitted from the plot. The continuous and the dashed lines show the elliptical and the S0 CMR zero point, respectively. Most of the blue S0 lie close to the cluster center, not on the outskirts.  {\label{colors}}}
\end{figure*}

The quality of the ACS data allows us to discern
merging activity among the cluster galaxies.  If we assume
that galaxy pairs with projected separations less than $20
h^{-1}_{70}$~kpc are physically associated, we observe 13 associated
early--type galaxies, nine galaxies of which lie on a filamentary
structure about $\approx$~100~kpc from the cluster center
(Fig.~\ref{cluster}).  As noted above, RXJ~0910+5422 lacks any
cD galaxy near the center of the X--ray emission 
(see also Fig.~\ref{cluster}), but rather has a filamentary
group of galaxies around the X--ray center. 

The nine early-type interacting galaxies
within this filamentary structure (at radius of
$\sim100 h^{-1}_{70}$~kpc from the X--ray center) include 3 unique pairs
(yellow arrows in Fig.~\ref{cluster}), plus a galaxy triplet
(the three components marked with red arrows in Fig.~\ref{cluster}).
Each of the 3 pairs consists of a bright elliptical with a smaller
companion (all closer than  $10 h^{-1}_{70}$~kpc), while the triplet is a large E with two smaller S0s
(also closer than  $10 h^{-1}_{70}$~kpc; one of these two S0s is ACS ID 1621, the S0 with an inward blue gradient). 
One of the pairs (the middle pair in the figure) and the two nearest
galaxies in the triplet, have essentially zero relative velocity and
thus are likely merger candidates.  
Also two of the ellipticals with blue inward gradients lie on the central filaments, and they are both small satellites of a larger galaxy.

The other two pairs, which do not lie on the filamentary structure, 
 are at ~$250h^{-1}_{70}$~kpc (two E with similar size, both with weak O~II
emission (Stanford et al. 2002) and ~$300 h^{-1}_{70}$~kpc (one S0/a
and one E of similar size) from the X--ray center, and have relative
velocities of  ~10,000~km/s and ~2000~km/s, respectively.  

The presence of the low-velocity pairs is consistent with the
low velocity dispersion, and provides evidence for the
on--going hierarchical growth of the cluster (e.g., van Dokkum
et al. 1999).
The pairing of the bright ellipticals with smaller S0s also argues against
the view that we are observing a blue S0-dominated group infalling 
into a red cluster elliptical population, but rather complex stellar
population evolution within a filamentary structure.
The red--sequence S0/a confirmed members also lie within this
filamentary structure.

The observations of a significant number of red galaxy pairs in a 
cluster at z~$\sim$~1 is interesting in the context of the
recent findings by van Dokkum (2005) of red galaxy interactions
in $\approx$~70\% of 86 early--type galaxies in a selected sample of nearby red galaxies from the MUSYC (Multiwavelength Survey by Yale--Chile; Gawiser et al. 2005) and the NOAO Deep Wide--Field Survey. 
This work concluded that most of the ellipticals in local samples were
assembled by red galaxy--galaxy mergers, denominated {\it dry} mergers because they would involve
gas--poor early--type galaxies.
At higher redshift, Tran et al. (2005b) confirmed red galaxy mergers first observed by van Dokkum (1999) in MS~1054-03 at z=0.83. Tran et al. selected mergers as associated pairs with projected separation less than $10 h^{-1}_{70}$~kpc and relative line--of--sight velocities less than 165~km/s. As in RXJ~0910+5422, the red early--type galaxies involved in these mergers are among the brightest cluster members. Their results suggest that most early--type galaxies grew from passive red galaxy--galaxy mergers.
In our sample we observe a triplet and three red galaxy pairs with projected distances less than $10 h^{-1}_{70}$~kpc. Of those, the triplet and one galaxy pair show zero relative velocity. Of the two other pairs, composed of a bright and a fainter companion, redshifts are not available for the faint companions.

\section{Cluster luminosity function}

To obtain a deeper understanding of the RXJ~0910+5422 galaxy population,
we constructed galaxy luminosity functions in the following way.  We start with the original
Sextractor catalog (described in Section~3).  All objects with
magnitudes brighter than $i_{F775W} = 21.1$~mag were considered as
foreground objects. Nine of these bright objects are confirmed 
non-members.  The remaining seven are objects that do not belong to the
red sequence and whose sizes and luminosities are much larger
than those of the confirmed members, in particular those of the 
bright red sequence galaxies; therefore they are very
unlikely to be at the cluster redshift.  

The contribution to the luminosity function from both foreground and
background field galaxies (hereinafter, the field) has been estimated from the galaxy counts in a
reference field. The control region is taken from the GOODS-S  
(Great Observatories
Origins Deep Survey--South; Giavalisco et al. 2004) ACS field,
observed in the same filter as the cluster field.  Point--like
objects were eliminated in a consistent way in the cluster and in the
control field, by identification of the stellar locus in the
diagnostic plot of the SExtractor parameters MAG\_AUTO vs FLUX\_RADIUS
(the selected objects have FWHM equal to the PSF in the image).
Cluster and control field luminosity functions were normalized to the
cluster area. Both cluster and field counts were binned with a bin
size of 0.5~mag.  For each bin, the field counts are subtracted from
the cluster counts, taking into account the extensive spectroscopic
sample (more than 60\% of the objects used for the LF determination
brighter than M$^{*}$ have measured redshifts).  Known interlopers
were excluded from the analysis.  The uncertainties in the cluster
counts after subtraction of the field contribution are calculated by adding in quadrature
Poissonian uncertainties.

\begin{figure*}
\centerline{\includegraphics{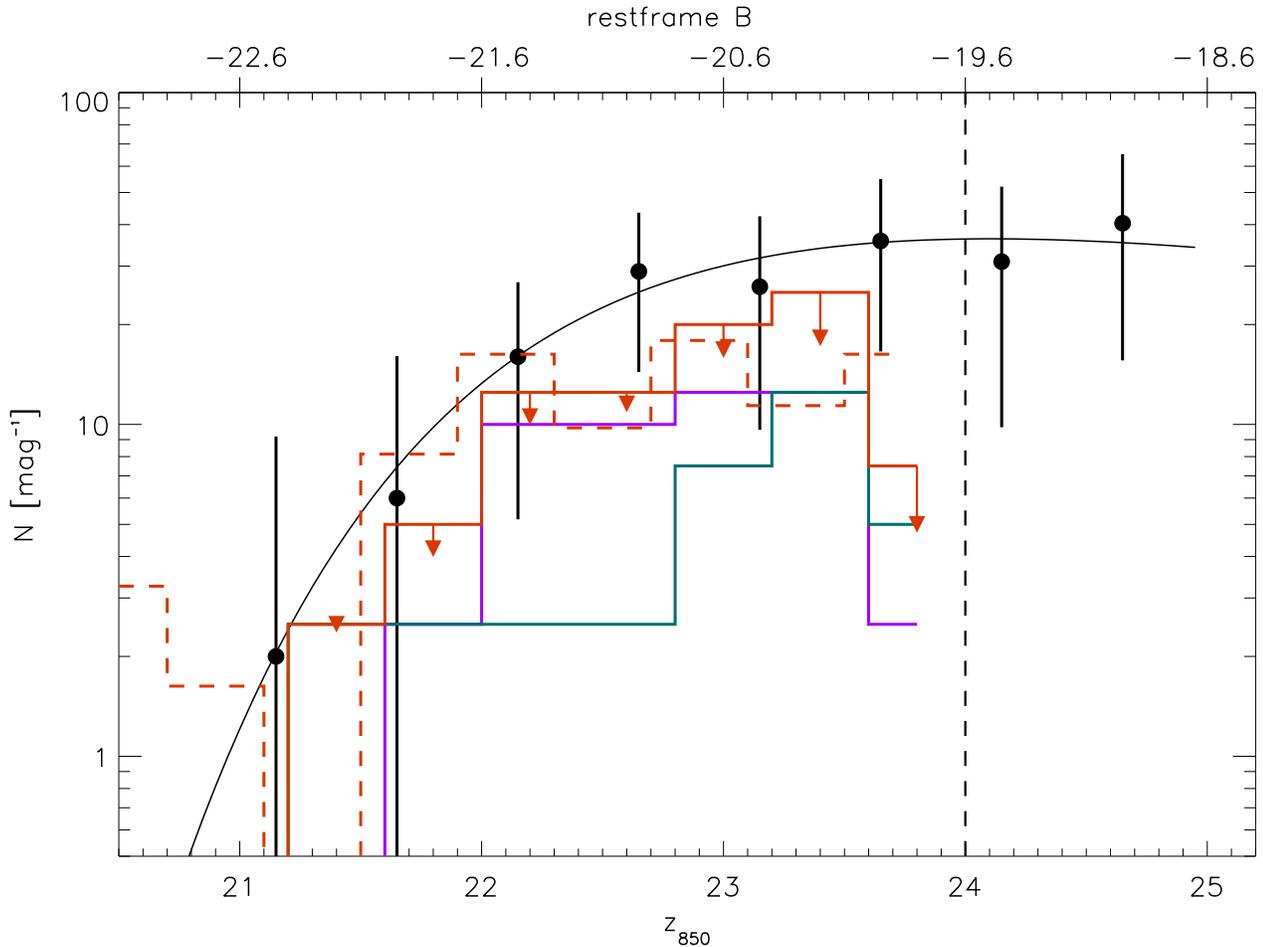}}
\caption {The $z_{850}$ luminosity functions for all galaxies (black circles with errors), 
as well as red sequence early-type (red), elliptical (blue), and S0+S0/a (green) galaxies are shown. 
The faint end of the red sequence luminosity function is dominated by S0s and S0/as. The solid line is the fit to the total luminosity function. We obtain M$^{*}=22.6^{+0.6}_{-0.7}$~mag and $\alpha=-0.75\pm0.4$.
 Galaxies have been classified morphologically down to $z_{850}=24$~mag. The histogram of red sequence galaxies in RDS1252.9-292 is shown as the dashed red line, as a comparison. The red arrows show the histogram values
after background subtraction. {\label{lumfun}}}
\end{figure*}

The luminosity functions are shown in Fig~\ref{lumfun}.
The filled circles with errors are the total background--corrected cluster
luminosity function.  We do not include errors from cosmic variance due to
the choice of the background control region. The red sequence elliptical and
spheroidal (S0 and S0/a) luminosity functions are shown respectively in
blue and green. 
The red histogram is the luminosity function of all early--type
galaxies with color $0.8 <(i_{775} - z_{850}) < 1.1$~mag, excluding
confirmed non--members. The histogram of red sequence galaxies in
RXJ1252.9-292 is shown as the dashed red line. The background contribution is very small
for the early-type sample. The red arrows show the histogram values
after background subtraction.
The rest-frame $B$~magnitudes are shown on the top of the plot,
calculated from colors obtained from the BC03
stellar population model and templates (Sbc, Scd) from Coleman, Wu \&
Weedman (1980). The solid black line is a Schechter function fit
to the total cluster luminosity function.  It is obtained by
calculating the C (Cash 1979) statistic (a 
maximum likelihood statistic to fit data with Poissonian errors)
 on a grid in the M$^{*}$ --
$\alpha$ plane for each combination of M$^{*}$ and $\alpha$: first the
normalization ($\phi^{*}$) is calculated in order to reproduce the
observed number of galaxies in the observed magnitude range, then the
C statistic is computed as $C=-2\Sigma_{i_{bin}} n_{i} \ln m_{i} -
m_{i} - \ln n_{i}!$, where $n_{i}$ is the observed number of galaxies
in the $i\,$th bin and $m_{i}$ is the number of galaxies predicted in
that bin by the Schechter function with parameters M$^{*}$, $\alpha$,
and $\phi^{*}$.

The combination M$^{*}$,$\alpha$ which minimizes the C statistic is
taken as the best-fit. If the C statistic is defined as
above, the 1- 2- and 3-$\sigma$ confidence levels for M$^{*}$ and $\alpha$
can be estimated from $\Delta C = 2.3, 6.17,11.8$. We obtain 
M$^{*}=22.6^{+0.6}_{-0.7}$~mag and $\alpha=-0.75\pm0.4$.

Most of the faint--end population is composed of S0 and S0/a galaxies. 
The two brightest galaxies in the red sequence are S0.
With respect to RXJ1252.9-292, a bright population of
red sequence ellipticals is missing in RXJ~0910+5422.
However, the large Poissonian errors on the bright end of the 
cluster population prevent us from definitively excluding the hypothesis
that
the two clusters could be drawn from the same parent population.
Similarly, small number statistics do not permit us to study the
luminosity functions of 
the different red and blue faint populations in this cluster.

\section{Discussion and Conclusions}

In this paper we have studied the color--magnitude relations of
galaxies in RXJ\,0910+5422 to constrain their ages and formation histories.
Our results show that the  color--magnitude relation for the elliptical
galaxies is consistent both in slope and scatter with that of RXJ\,1252.9-292 
(Blakeslee et al. 2003a, Lidman et al. 2004) and recent results from Holden et
al. (2004) and De Lucia et al. (2004), confirming that elliptical
galaxies in galaxy clusters show a universal color--magnitude relation
consistent with an old passively evolving population even at $z \sim 1$. 
From the color--magnitude relation of the ellipticals, we derive a mean
luminosity--weighted age $\overline t > 3.3$~Gyr
($z_f > $~3). 
 
We find that the S0s in RXJ~0910+5422 define a color--magnitude sequence
with a scatter similar to that found for the ellipticals, but shifted bluer
by  $0.07 \pm 0.02$~mag.   This is peculiar with respect to previous cluster
studies, which more typically found that the S0s followed the same CMR
as the ellipticals, but with somewhat larger scatter (Bower et al. 1992; Ellis et al.\ 1997; Stanford et al.\ 1997; L\'opez--Cruz et al. 2004; Stanford, Eisenhardt, \& Dickinson 1998; van Dokkum et al. 2000, 2001; Blakeslee et al. 2003a; Holden et al. 2004). Only one earlier study, van Dokkum et al. 
(1998) found a significantly bluer S0 population. 
We examine this
population of blue S0s in some detail, noting that there is a strong
predominance of flattened systems with axial ratios $\frac{b}{a} > 0.7$,
and conclude that the face-on members of the population have likely been
classified as ellipticals.  If so, the color offset between the two
classes would become even more significant, and the true CMR scatter
for the ellipticals would be slightly lower than we have estimated.
 This peculiarity is not observed in other clusters of our ACS
 Intermediate 
Redshift 
Cluster Survey, and its amplitude is smaller 
than the uncertainties adopted in Postman et al. (2005).

If the observed color difference between the ellipticals and S0s is mainly
due to metallicity at the same age, this would imply that the redder 
ellipticals
were able to retain more metals than the S0s, i.e., they are more massive. 
However, current data suggest that the CMR is mainly the result of a metallicity--mass
(i.e. metallicity--magnitude) relation 
(e.g., Kodama \& Arimoto 1997; Kauffman \& Charlot 1998
Vazdekis et al. 2001; Bernardi et al. 2005).
This implies that we do
not expect large metallicity variations at a given magnitude.
If, instead, the offset is mainly due to age, then the implied age difference would be
$\sim$1~Gyr for single-burst solar metallicity BC03 models.  It could
also result from different star formation histories, with the S0s experiencing
a more extended period of star formation.  A model with solar metallicity
and with an exponential decay of the star formation reproduces the offset
at a galaxy mean age of $\approx$~3.5~Gyr.

The blue S0s may comprise a group infalling from the field onto a 
more evolved red cluster population, or they may be
a transitional cluster population not yet evolved all the way onto the
elliptical red sequence (van Dokkum \& Franx 2001). Assuming
passive evolution, they will reach this red sequence after about 1~Gyr.
High fractions of faint blue late type galaxies were observed in 
substructures infalling in a main cluster (e.g. Abraham et al. 1996; Tran et al. 2005a), and proposed as the progenitors of faint S0s in clusters.
The view of this
cluster as a structure still in formation is supported by X--ray
observations of the cluster temperature structure (Stanford et al. 2002), 
the lack of a cD galaxy, and its
filamentary structure that suggests merging of substructures. However,
we derive a small cluster velocity dispersion, unusual for merging
substructures.  
Moreover, the blue S0s in this sample span the same
luminosity range of the bright ellipticals, are distributed towards the center of the
cluster, and some of the faintest ones are physically associated with
brighter ellipticals belonging to the central filamentary structure.
These elements would argue against the bluer S0s being a young group merging with an existing red cluster population, and  support the hypothesis that we are observing a transitional blue S0 population in a cluster core that 
is still evolving onto the elliptical red sequence. 
This result is also consistent with the deficit of S0s observed in our ACS cluster sample, when compared to lower 
redshift samples, that implies that the S0 population is  not
yet in place but still forming in clusters at redshifts around unity (Postman et al. 2005).
Interestingly, we also observe in this cluster potential progenitors for bright S0 galaxies: four bright spirals (spectroscopically confirmed cluster members)
with $z_{850}$ brighter than 22.5~mag and $(i_{775}-z_{850})$
between 0.5 and 1.3~mag. 

Red galaxy pairs are also observed. A triplet and three red galaxy pairs have projected distances less than $10 h^{-1}_{70}$~kpc, and of those, the triplet and one pair show zero relative velocity. This would be the evidence of red galaxy mergers at z$\sim$~1. van Dokkum (2005) and Tran et al. (2005b) have observed mergers of red galaxies in a nearby elliptical sample and in  MS 1054-03 at z=0.83, respectively. They suggested a scenario in which most of the early--type galaxies were formed from passive red galaxy--galaxy mergers, called {\it dry} mergers, because they involve gas--poor early--type galaxies.

Future papers will analyze the ages and masses of the cluster members
using our optical spectroscopy along with newly obtained Spitzer IRAC imaging.
A larger sample would be needed to draw firmer conclusions about the formation
of S0s.

\begin{acknowledgements}
ACS was developed under NASA contract NAS 5-32865, and this research 
has been supported by NASA grant NAG5-7697 and 
by an equipment grant from  Sun Microsystems, Inc.  
The {Space Telescope Science
Institute} is operated by AURA Inc., under NASA contract NAS5-26555.
We are grateful to K.~Anderson, J.~McCann, S.~Busching, A.~Framarini, S.~Barkhouser,
and T.~Allen for their invaluable contributions to the ACS project at JHU. 
We thank W. J. McCann for the use of the FITSCUT routine for our color images.
SM thanks Tadayuki Kodama for useful discussions.
\end{acknowledgements}

\newpage


\clearpage

\begin{small}
\begin{table*}
\begin{center}
\caption{Color--Magnitude Relations \label{results}}
\vspace{0.25cm}
\begin{tabular}{cccccccccccc}
\tableline \tableline\\
 Sample &$N$&$c_0$&$Slope$ & $\sigma_{int}$\\
&&(mag)&&(mag)&\\
\tableline \tableline\\

E+S0+S0/a$^1$&          31&      0.99 $\pm$    0.01&   -0.030 $\pm$
    0.020&    0.060 $\pm$   0.008\\
E+S0$^{1a}$&          14&      1.00 $\pm$    0.02&   -0.010 $\pm$
    0.033&    0.054 $\pm$   0.009\\
E$^1$&          19&      1.02 $\pm$    0.01&   -0.033 $\pm$
    0.015&    0.042 $\pm$    0.011\\
E$^{1a}$&          10&      1.02 $\pm$    0.04&   -0.020 $\pm$
    0.044&    0.047 $\pm$    0.022\\
S0$^1$&           9&     0.95 $\pm$    0.02&   0.005 $\pm$
    0.023&    0.044 $\pm$    0.02\\
S0+S0/a$^{1}$&          12&     0.95 $\pm$    0.02&  -0.007 $\pm$
    0.027&    0.057 $\pm$    0.015\\  \tableline \\

E+S0+S0/a$^2$&          32&      0.99 $\pm$    0.01&   -0.032 $\pm$
    0.019&    0.060 $\pm$   0.008\\
E+S0$^{2a}$&          15&      0.99 $\pm$    0.02&   -0.021 $\pm$
    0.034&    0.054 $\pm$   0.009\\
E$^2$&          19&      1.02 $\pm$    0.01&   -0.033 $\pm$
    0.015&    0.042 $\pm$    0.011\\
E$^{2a}$&          10&      1.02 $\pm$    0.04&   -0.020 $\pm$
    0.044&    0.047 $\pm$    0.022\\
S0$^2$&          10&     0.95 $\pm$    0.02&   -0.012 $\pm$
    0.036&    0.051 $\pm$    0.018\\
S0+S0/a$^{2}$&          13&     0.96 $\pm$    0.02&   -0.015 $\pm$
    0.033&    0.065 $\pm$    0.015\\ \tableline \\

E+S0+S0/a$^3$&          34&   0.99    $\pm$    0.01&   -0.036 $\pm$
    0.018&    0.059 $\pm$   0.008\\
E+S0$^{3a}$&          15&     0.99 $\pm$    0.02&   -0.022 $\pm$
    0.035&    0.054 $\pm$   0.009\\
E$^3$&          20&      1.01 $\pm$    0.01&   -0.032 $\pm$
    0.015&    0.044 $\pm$    0.010\\
E$^{3a}$&          10&      1.02 $\pm$    0.04&   -0.021 $\pm$
    0.046&    0.047 $\pm$    0.022\\
S0$^3$&          11&     0.96 $\pm$    0.02&   -0.022 $\pm$
    0.038&    0.053 $\pm$    0.015\\
S0+S0/a$^{3}$&          14&     0.96 $\pm$    0.02&   -0.024 $\pm$
    0.034&    0.065 $\pm$    0.013\\

\tableline \tableline
\end{tabular}
\end{center}
$1$: within 1~$\arcmin$ \\
$2$: within 1.5~$\arcmin$ \\
$3$: within 2~$\arcmin$ \\
$a$: only confirmed members
\end{table*}
\end{small}


\end{document}